\documentclass[reprint,
superscriptaddress,
 amsmath,amssymb,
 aps,
pra,
]{revtex4-2}

\usepackage{graphicx}
\usepackage{dcolumn}
\usepackage{bm}
\usepackage[utf8]{inputenc}
\usepackage{subcaption}
\usepackage{mhchem}
\usepackage{multirow}
\usepackage{makecell}
\usepackage{hyperref}
\usepackage{color,soul}
\usepackage{chemformula}
\usepackage{comment}
\usepackage{textcmds}
\usepackage{caption}
\usepackage{chemscheme}
\usepackage{amsfonts}
\usepackage{csquotes}
\usepackage{adjustbox}
\usepackage{booktabs}
\usepackage{algorithm}
\usepackage{algpseudocode}
\usepackage{nicefrac}

\begin{document}

\preprint{APS/123-QED}

\title{A Critical Examination of Active Learning Workflows in Materials Science}

\author{Akhil S. Nair}
\email{akhil.sugathan.nair@fu-berlin.de}
\affiliation{ The NOMAD Laboratory at the Fritz Haber Institute of the Max Planck Society, Faradayweg 4-6, D-14195 Berlin, Germany}
\affiliation{Institut für Chemie und Biochemie, Freie Universität Berlin, Arnimallee 22, 14195, Berlin, Germany}
\author{Lucas Foppa}
\affiliation{ The NOMAD Laboratory at the Fritz Haber Institute of the Max Planck Society, Faradayweg 4-6, D-14195 Berlin, Germany}


\date{\today}

\begin{abstract}
Active learning (AL) plays a critical role in materials science, enabling applications such as the construction of machine-learning interatomic potentials for atomistic simulations and the operation of self-driving laboratories. Despite its widespread use, the reliability and effectiveness of AL workflows depend on implicit design assumptions that are rarely examined systematically. Here, we critically assess AL workflows deployed in materials science and investigate how key design choices, such as surrogate models, sampling strategies, uncertainty quantification and evaluation metrics, relate to their performance. By identifying common pitfalls and discussing practical mitigation strategies, we provide guidance to practitioners for the efficient design, assessment, and interpretation of AL workflows in materials science.
\end{abstract}

\maketitle

\section{Introduction}
The application of machine learning (ML) in materials science and engineering has expanded rapidly, enabling progress across a wide range of tasks, including high-throughput screening, accelerated materials property prediction and autonomous experimentation \cite{butler2018machine, vasudevan2021machine, axelrod2022learning, sumpter2015bridge}. Central to many of these efforts is the challenge of learning complex mappings between structure, composition, processing conditions, and materials properties, which are typically non-linear and only partially understood. By learning directly from data rather than relying on explicit physical models, ML methods provide a flexible framework for capturing such complex relationships and have demonstrated transformative potential in materials science \cite{pyzer2022accelerating,merchant2023scaling}. However, their success is fundamentally constrained by the availability, accuracy and quality of the training data. Despite advances in high-throughput experimentation and simulation frameworks \cite{maier2007combinatorial, curtarolo2013high}, generating consistent, high-fidelity materials data remains time- and resource-intensive. As a result, many materials-science applications operate in regimes better characterized as data-scarce rather than data-rich \cite{xu2023small}. Moreover, materials datasets are often not \qq{statistically representative}, i.e., they tend to be shaped by prior domain knowledge, biased curation strategies, or feasibility constraints, and frequently fail to adequately represent the broader space of relevant structures, compositions, or other relevant parameters. As a result, ML models trained on such data are prone to unreliable performance when applied to unexplored or underrepresented regions of the data space \cite{bauer2024roadmap,karande2022strategic}.

\vspace{4pt}
Active machine learning (hereafter referred to as active learning, AL) has emerged as a promising framework for addressing these challenges in ML-driven materials science. As formalized by Cohn et al. \cite{cohn1996active}, \qq{AL studies the closed-loop phenomenon of a learner selecting actions or making queries that influence what data are added to its training set}. By employing adaptive sampling strategies, AL prioritizes the acquisition of critical  data points \cite{cohn1994improving}. AL is often conceptualized as an efficient labeling strategy. Its applications in materials science are broad, ranging from training data generation \cite{gubaev2018machine, jinnouchi2020fly} to guiding closed-loop experimentation and autonomous materials discovery \cite{nie2024active, suvarna2024active}. To illustrate the diversity of AL use cases, we briefly highlight two representative application settings that are currently of significant interest: (i) ML-driven atomistic simulations using machine-learning interatomic potentials (MLIPs) and (ii) optimization-based workflows for materials discovery. In the context of MLIPs, a central challenge is capturing rare but physically important phenomena such as bond breaking, phase transitions, or defect migration that occur infrequently, which are often poorly represented in training datasets. Here, AL can be used to ensure the inclusion of underrepresented yet physically essential configurations, particularly those associated with rare events or extreme conditions encountered during MLIP-guided molecular dynamics (MD) simulations \cite{kang2024accelerating, smith2018less, verdi2021thermal}. In autonomous laboratories focused on materials property optimization, the primary difficulty lies in identifying statistically exceptional materials within vast design spaces (discussed in detail later). In this setting, AL aids in achieving the goal of maximizing the likelihood of selecting promising candidates for evaluation while minimizing the cost required to find materials with certain properties. Despite these differing objectives, AL is typically implemented as a workflow composed of interwined design choices, such as surrogate models, sampling strategies, uncertainty quantification and performance evaluation metrics. 

While AL workflows are now routinely applied in materials science \cite{lookman2019active, bassman2018active, ozbayram2025batch}, their design choices vary widely, even for closely related tasks. This diversity partly reflects the aforementioned breadth of materials science applications and is therefore neither surprising nor inherently problematic. However, it also introduces substantial inconsistency in how AL workflows are designed, executed, and assessed, making it difficult to compare performance or draw general conclusions across studies. For example, in data generation tasks, AL workflows may be designed to prioritize unfamiliar data points (e.g. novel compositions or structures) \cite{hu2024designing}, reduce biases in the initial dataset \cite{zhang2023entropy}, improve predictive performance of ML models \cite{jose2024informative}, or decrease model uncertainty \cite{kang2024accelerating}. While these objectives are often interrelated, practitioners typically evaluate the effectiveness of AL using only a subset of metrics, most commonly improvements in model accuracy, without systematically assessing whether coverage of critical regions of the data space has also improved. Similarly, in materials discovery applications, the choice of surrogate models \cite{bauer2024roadmap}, acquisition functions (vida infra) \cite{wang2022benchmarking}, and uncertainty quantification methods \cite{tian2020role} can strongly influence outcomes, yet these choices are often made heuristically and evaluated using application-specific criteria that hinder cross-study comparison. Moreover, a fundamental question remains insufficiently addressed: \qq{to what extent does AL provide benefits beyond those achievable through simpler data selection strategies based on human intuition, prior domain knowledge, or hand-crafted rules?.} In other words, it is often unclear whether observed performance gains stem from a principled, algorithmic AL framework or could instead be achieved by informed, human-guided selection of data points without an explicit AL loop. These issues highlight the need for modular and standardized AL workflows that enable consistent evaluation and comparison across applications, while still allowing flexibility to accommodate domain-specific objectives. Without such structure, AL workflows risk being inefficient, difficult to interpret, or misdirected, potentially leading to unnecessary computational or experimental cost and convergence toward suboptimal solutions.

In this perspective, we critically examine AL workflows commonly employed in materials science, focusing on two key applications of data generation and materials discovery. We begin with a brief overview of the AL methodology and analyze the strengths and limitations of tools and techniques used at different stages of AL workflows. We do not aim to provide a comprehensive review here and instead refer readers to existing review articles for the broader context \cite{lookman2019active, kulichenko2024data}. Our focus is to raise awareness of practitioner-driven biases in the design choices that can impact the performance of AL workflows. Building on this analysis, we propose guidelines to support the rigorous design, assessment, and interpretation of AL workflows in materials science. 
\section{Active Learning Methodology}
An AL algorithm involves performing data generation iteratively and adaptively, with the goal of selecting the most important data points for labelling under a limited evaluation budget. In materials science, the \qq{labels} are typically obtained from an oracle such as a high-fidelity simulation, a physical experiment, or expert annotation (Fig. 1a), all of which are costly in terms of time or resources. An AL workflow therefore seeks to maximize learning efficiency by prioritizing which data points to evaluate next, rather than relying on sampling performed randomly \cite{konyushkova2017learning}. Because AL algorithms often construct a sequence of targeted queries, it is sometimes referred to in the literature as \qq{query learning} \cite{campbell2000query,hwang1991query} or\qq{sequential learning} \cite{rohr2020benchmarking,khosravi2024data}. Several AL variants exist depending on how unlabeled data are accessed or generated, including query synthesis, where new candidate inputs are generated by the learner \cite{angluin1988queries}, and stream-based AL, where data arrive sequentially and must be queried or discarded on the fly \cite{atlas1989training}. However, we will use the term \qq{active learning} and stick to discussing pool-based AL \cite{lewis1995sequential} where it is assumed that a relatively larger pool of unlabelled data is available for labelling, because of its simplicity and its widely adopted usage in the field.  The objective of an AL algorithm is to achieve a task-specific goal (e.g. improve model performance) while minimizing the overall cost of labeling (or another user-defined criterion) through strategic sampling (Algorithm 1). In a supervised learning setting, let \( \mathcal{D}_{\text{init}} \subset \mathbb{X} \times \mathbb{Y} \) denote the initially available labelled dataset (hereafter referred to as \qq{seed data}), and let \( \mathcal{U} \subset \mathbb{X} \) represent the pool of unlabeled data (hereafter referred to as \qq{pool data}). Here, \(\mathbb{X} \) represents the input space (e.g. crystal structures, compositions) and \(\mathbb{Y} \) represents the corresponding outputs of interest (e.g. energies, forces, or materials properties). A surrogate ML model \( M \) is trained to approximate an unknown target function \( f: \mathbb{X} \to \mathbb{Y} \), mapping \( x \in \mathbb{X} \) to \( y \in \mathbb{Y} \).  At each iteration, a sampling strategy \( Q \) evaluates the pool data using the current model and selects one or more candidate inputs \( x^* \in \mathcal{U} \), deemed most informative according to a predefined criterion. These candidates are then evaluated by an oracle \( \mathcal{O} \) to obtain labels \( y^* = \mathcal{O}(x^*) \). after which the labeled dataset and unlabeled pool are updated:
\vspace{-4pt}
\[
\mathcal{D}_{\text{init}} \leftarrow \mathcal{D}_{\text{init}} \cup \{(x^*, y^*)\}, \quad \mathcal{U} \leftarrow \mathcal{U} \setminus \{x^*\}
\]

This process is repeated until a stopping criterion is satisfied, such as a performance threshold or exhaustion of a query budget \( B \in \mathbb{N} \).

\begin{algorithm}[H]
\caption{Pool-Based Active Learning Algorithm}\label{algorithm:1}
\begin{algorithmic}[1]
\Require Initial labeled dataset \( \mathcal{D}_{\text{init}} \subset \mathbb{X} \times \mathbb{Y} \), unlabeled pool \( \mathcal{U} \subset \mathbb{X} \), oracle \( \mathcal{O} \), sampling strategy \( Q \), budget \( B \in \mathbb{N} \), surrogate model \( M \)
\Ensure Trained model \( M \) approximating \( f: \mathbb{X} \to \mathbb{Y} \)

\State Train surrogate model \( M \) on \( \mathcal{D}_{\text{init}} \)
\While{stopping criterion not met \textbf{and} \( B > 0 \)}
    \State Select query point \( x^* \gets Q(M, \mathcal{U}) \)
    \State Obtain label \( y^* \gets \mathcal{O}(x^*) \)
    \State Update dataset \( \mathcal{D}_{\text{init}} \leftarrow \mathcal{D}_{\text{init}} \cup \{(x^*, y^*)\} \)
    \State Remove queried sample \( \mathcal{U} \leftarrow \mathcal{U} \setminus \{x^*\} \)
    \State Retrain \( M \) on updated \( \mathcal{D}_{\text{init}} \)
    \State \( B \leftarrow B - 1 \)
\EndWhile
\State \Return Trained model \( M \)
\end{algorithmic}
\end{algorithm}

Since the AL algorithm involves weighing of unlabeled data and strategic selection of the best samples for labelling, it offers a targeted approach that allows the model to learn efficiently from fewer samples, which is critical for a data-limited domain like materials science. More generally, AL provides a data-efficient framework for exploring high-dimensional \textit{design spaces}, i.e., the set of candidate inputs under consideration for a given task (e.g. possible alloy compositions). By integrating model training and data generation within a single closed-loop workflow, AL supports targeted exploration of these spaces without resorting to exhaustive enumeration, thereby enabling the identification of promising materials or configurations at reduced computational or experimental cost. The iterative nature of AL further allows models to adapt as new data are acquired, reducing reliance on initial, potentially biased datasets  as learning progresses.

\begin{figure*}[ht]
    \centering
    \includegraphics[width=\linewidth]{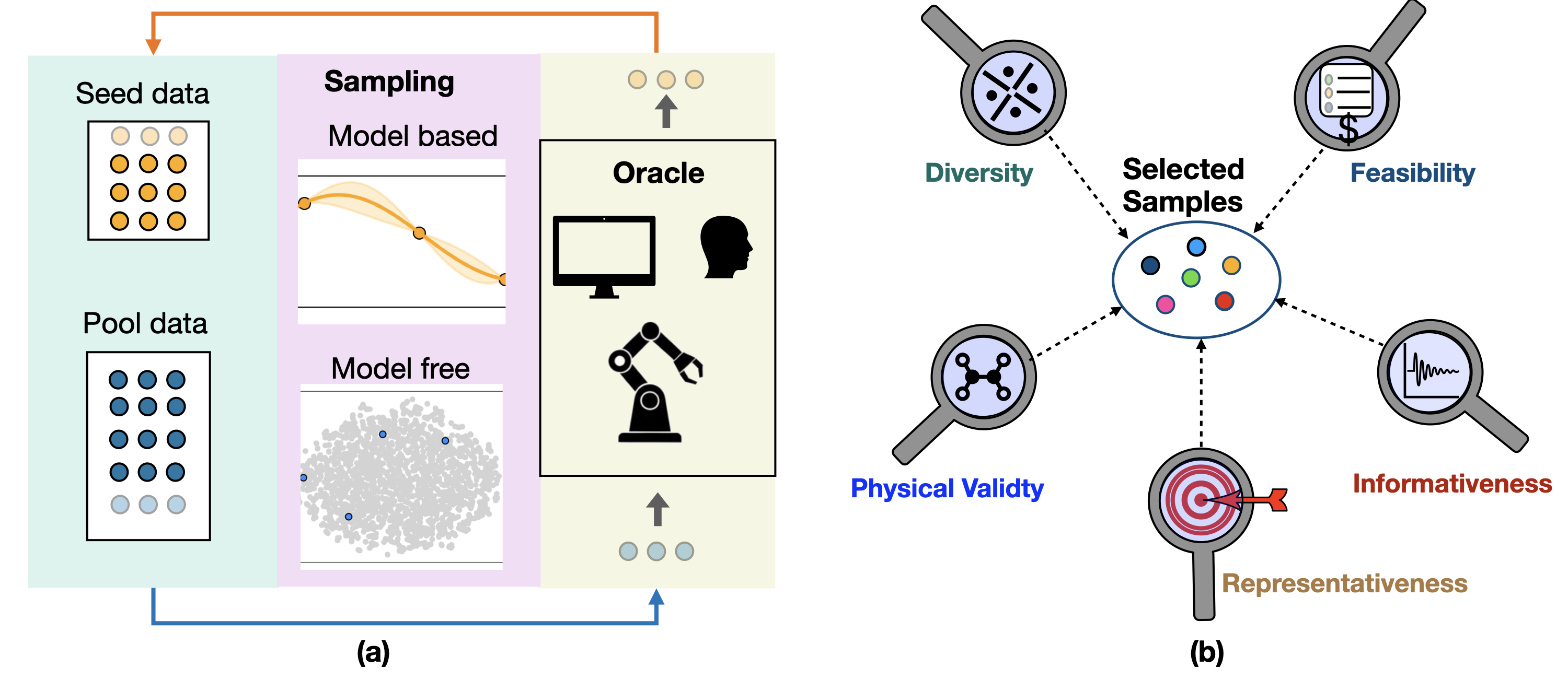}
    \captionsetup{justification=justified}
    \caption{Schematic representation of (a) an AL workflow where both model-based and model-free AL strategies can be employed to acquire samples from pool data and update the seed data by interacting with an oracle, (b) various factors that need to be considered while designing the sampling strategy for AL.}
    \label{Fig:1}
\end{figure*}

\vspace{4pt}

\section{Active Learning Applications and Challenges}
In this section, we highlight the key challenges in employing AL workflows for materials science problems through two representative application domains: (i) materials data generation and (ii) materials discovery. It is to be noted that here, the term \qq{materials discovery} refers broadly to identifying materials with desirable properties, whether among previously synthesized, computationally generated, or yet-to-be-synthesized candidates. In this sense, discovery entails jointly establishing material identity and properties within a vast, largely unexplored materials space. For both i) and ii), we discuss, based on existing literature, how AL can be useful and identify critical challenges that remain unaddressed or are frequently overlooked. We provide practical guidelines and recommendations for addressing these challenges in future AL-driven materials research in the Outlook section.

\subsection{Data Generation/Redundancy Mitigation}
\textbf{Redundancy problem in materials data}: Materials datasets are often curated based on prior knowledge (e.g. well-known materials with desirable properties) or ease of access (e.g. from existing data repositories). These practices can introduce substantial redundancy and lead to biased coverage of the broader materials space. Recent work by Li et al. \cite{li2023exploiting} demonstrated that widely used computational materials databases, including the Materials Project (MP) \cite{jain2013commentary} and OQMD \cite{kirklin2015open}, contain significant data redundancy. Using a data-pruning strategy, they showed that removing a large fraction of redundant entries neither degraded in-distribution model performance nor improved out-of-distribution (OOD) generalization for ML models trained on the full datasets (Fig. 2a).  While such redundancy is perhaps unsurprising given the high-throughput nature of computational database generation, biased sampling can persist even when subsets of these databases are selectively curated. For example, in many materials discovery studies \cite{merchant2023scaling,lyngby2022data}, candidate materials are preferentially chosen near the convex hull of a given compositional phase diagram, which can distort an ML model’s representation of the underlying stability landscape. Indeed, Bartel et al. \cite{bartel2020critical} showed that ML models can accurately predict formation energies yet still perform poorly in classifying stable versus unstable materials, particularly in sparsely sampled compositional spaces such as underrepresented quaternary systems in MP. In such scenarios, AL offers a promising alternative by adaptively focusing on data in the underexplored or informative regions of the materials space. By doing so, AL can enable the construction of smaller, more relevant datasets that mitigate redundancy while maintaining or even improving predictive performance.





\textbf{Inadequate sampling strategy :} One of the key components of an AL workflow is the \textit{sampling strategy} which is used to assign the aforementioned \textit{relevance} of samples from an unlabelled pool. Many of the AL workflows adapted in materials science \cite{jose2024informative,sheng2020active}, especially those employed for MLIP training \cite{bhatia2025leveraging,zhang2019active} uses \textit{informativeness}, i.e., ability of a sample to improve the model performance, as the sole sampling criterion. However, an effective AL sampling strategy is inherently multi-faceted and cannot be fully captured by informativeness alone (Fig. 1b). A key complementary criterion is \textit{representativeness}, which assesses whether a queried sample reflects the structure of the unlabeled data distribution \cite{settles2009active}.  This is particularly important to prevent sampling extreme outliers (e.g. highly distorted structures) that are not statistically representing the remaining, relevant materials space of interest. Various methods have been proposed by the ML community to include the representativeness factor \cite{wang2015querying,li2012active}, or jointly optimize informativeness and representativeness \cite{du2015exploring,huang2010active}, yet most AL workflows in materials science remain focused on informativeness due to the ease of monitoring via proxies such as model test errors. Beyond these standard criteria, additional materials-science specific factors need to be considered for sampling. These include: (i) \textit{diversity}, which ensures that selected samples are sufficiently distinct relative to both the seed data and between themselves to avoid redundancy \cite{zaverkin2022exploring,brinker2003incorporating} (ii) \textit{physical validity}, ensuring that samples are chemically and physically meaningful, for example by excluding unphysical MD configurations or crystal structures under extreme conditions and (iii) \textit{feasibility}, which accounts for practical constraints such as computational cost, favoring candidates that provide maximal information without incurring excessive expense (e.g. extremely large unit cells). In addition, for AL campaigns for which a reasonably large seed data are already available, pre-existing biases can persist, such as overrepresentation of certain chemistries, phases, or structural motifs. If these biases are not identified and addressed at the outset, AL may inadvertently reinforce them, leading to a sampling that further skews the dataset. Failure to consider these criteria while designing the sampling strategy can therefore severely limit the reliability of AL in materials science. 

\begin{figure*}[ht]
    \centering
    \includegraphics[width=\linewidth]{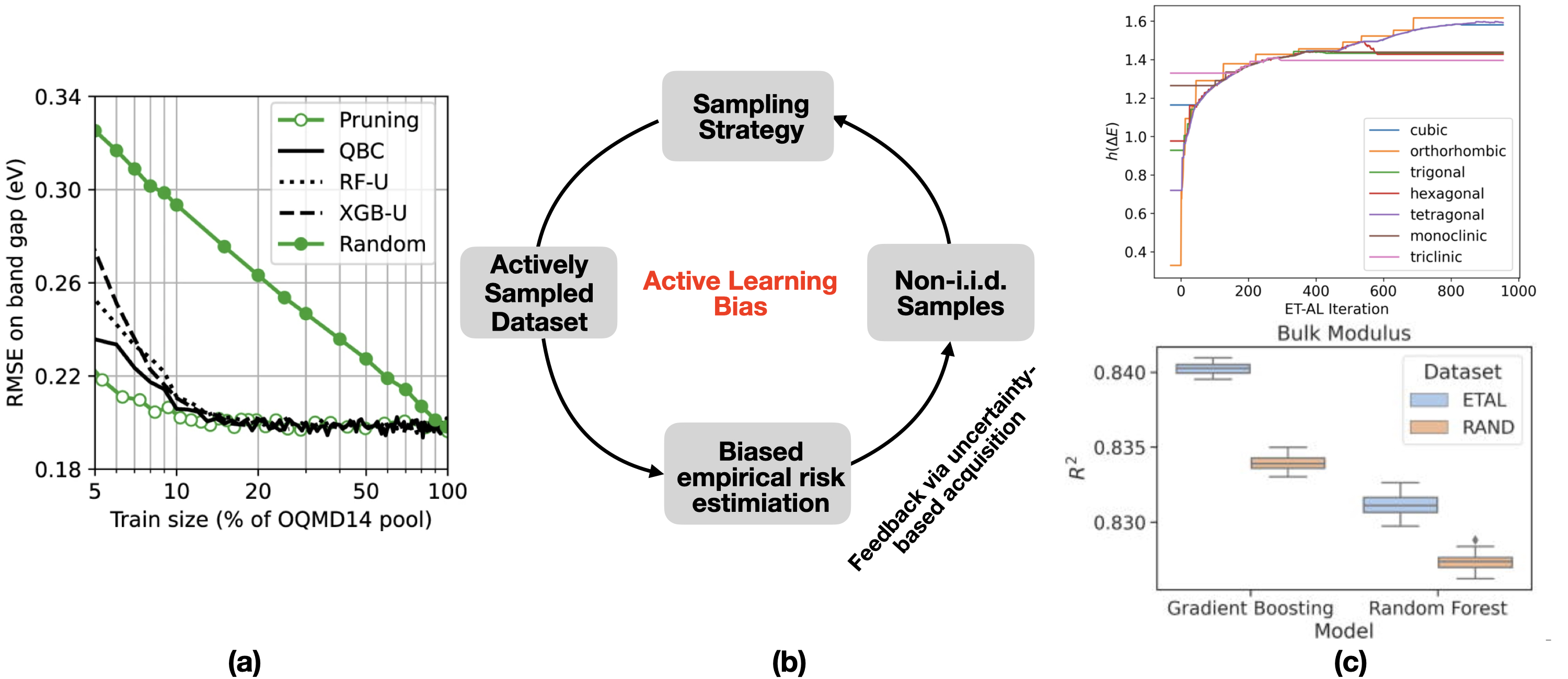}
    \captionsetup{justification=justified}
    \caption{(a) Active learning reduces redundancy in materials datasets: Performance of XGBoost (XGB) and Random Forests (RF) models on band gap prediction  trained on datasets obtained by uncertainty guided active learning, pruning, and random sampling from the OQMD14 dataset. Comparable accuracy is achieved with only 10\% of the data, highlighting substantial redundancy in the dataset, Adapted with permission from Ref. \cite{li2023exploiting}, Copyright 2023 Springer (b) schematic representation of active learning bias induced by sampling of data points do not following i.i.d assumption,  (c) information-entropy guided active learning (ETAL) minimizing the large structure-stability bias by improving the coverage of less symmetric crystal systems in the JARVIS classical force-field inspired descriptors dataset (top panel) and improved performance of ML models trained on such active learned dataset compared to random sampled dataset (bottom panel). Adapted with permission from Ref. \cite{zhang2023entropy}, Copyright 2023 American Institute of Physics.}
    \label{Fig:2}
\end{figure*}

\textbf{The ill-addressed active learning bias :} While AL can mitigate the redundancy problem, it can paradoxically introduce a new form of bias, termed as active learning bias (ALB) \cite{mackay1992information,dasgupta2011two}. This arises because during AL, samples are no longer drawn independently and identically distributed (i.i.d.), a fundamental assumption underlying ML model training.  As a result, actively curated training sets may deviate substantially from the application-relevant data distribution in the materials space of interest  (Fig. 2b). This deviation has important implications for model training and evaluation. Standard empirical risk minimization, which optimizes model parameters by minimizing the average loss over the training data, implicitly assumes that the training set is representative of the target distribution. When this assumption is violated, as is often the case in AL, model performance measured on finite or randomly curated test sets may reflect optimization with respect to a biased objective rather than genuine generalization across the materials domain. In materials science applications, this can manifest as models that perform well on actively sampled configurations while failing to generalize to unexplored chemistries or structures. Statistical corrections have been proposed to mitigate ALB, such as reweighting the training loss by the inverse probability of sample acquisition \cite{farquhar2021statistical}, thereby partially restoring consistency with the underlying data distribution. However, such approaches have not yet been systematically explored in materials science. Moreover, reweighting introduces additional subtleties: in overparameterized models, including deep neural networks, ALB can sometimes act as an implicit form of regularization, reducing overfitting and even improving apparent generalization \cite{murray2022addressing}. While this effect may be beneficial in practice, it complicates the interpretation of AL performance, as improvements may stem from sampling-induced regularization rather than principled coverage of the materials space. Because the magnitude and impact of ALB depend strongly on the mismatch between the seed dataset, the unlabeled pool, and the target application domain, careful attention to data distributions is essential. Incorporating explicit analysis of distributional coverage using tools such as dimensionality reduction \cite{hu2024realistic} or density estimation \cite{schultz2025general} can therefore provide critical context for evaluating AL outcomes and for designing more robust, application-aware AL workflows \cite{yang2018active,xiong2013active}. Without such considerations, AL strategies risk reinforcing hidden biases, limiting transferability, and overstating the effectiveness during the workflow deployment.

\begin{table*}[!ht]
\captionsetup[table]{position=top}
\caption{Examples of model-free active learning strategies and representative applications in materials science.}
\label{Table_ModelFreeAL}
\centering
\captionsetup{justification=raggedright, singlelinecheck=false}
\large
\begin{adjustbox}{width=1\textwidth}
\renewcommand{\arraystretch}{1.5}
\begin{tabular}{lccc}
\hline
\textbf{Sampling Criterion} & \textbf{Methods} & \textbf{Central Idea} & \textbf{Application in Materials Science} \\
\hline

\multirow{2}{*}{\textbf{Informativeness}} 
& Entropy-guided
& \begin{tabular}[c]{@{}p{6cm}@{}} 
Select samples that maximize information entropy computed from structural or descriptor distributions, enabling bias reduction and improved coverage without model uncertainty estimates.
\end{tabular}

& \begin{tabular}[c]{@{}p{6cm}@{}} 
Curation of bias-minimized crystal structure datasets \cite{zhang2023entropy}; 
Model-free uncertainty for MLIP-driven MD \cite{schwalbe2025model,pa2025information}
\end{tabular} \\
 
& Mutual-information-guided
& \begin{tabular}[c]{@{}p{6cm}@{}} 
Maximize mutual information between candidate samples and the existing dataset or experimental objectives to guide efficient discovery.
\end{tabular}
& \begin{tabular}[c]{@{}p{6cm}@{}} 
Phase-change memory material discovery \cite{kusne2020fly}
\end{tabular} 
\\
\hline

\multirow{2}{*}{\textbf{Representativeness}} 
& Clustering
& \begin{tabular}[c]{@{}p{6cm}@{}} 
Partition the unlabeled pool into clusters and select representative samples (e.g. cluster centroids) to preserve global coverage of the distribution and avoid oversampling statistical outliers.
\end{tabular}

& \begin{tabular}[c]{@{}p{6cm}@{}} 
Discovery of perovskite oxides for oxygen evolution catalysis \cite{moon2024active}; 
Training data generation for MLIPs \cite{qi2024robust}
\end{tabular} 
\\
& Density estimation  
& \begin{tabular}[c]{@{}p{6cm}@{}} 
Prioritize samples located in high-density regions of the unlabeled data distribution to ensure representativeness.
\end{tabular}

& \begin{tabular}[c]{@{}p{6cm}@{}} 
Functionalized nanoporous materials (MOFs/COFs) property prediction \cite{gkatsis2025density}; Assessing out-of-distribution performance of ML models \cite{schultz2025general}
\end{tabular} \\
\hline

\multirow{3}{*}{\textbf{Diversity}} 

& Distance-based
& \begin{tabular}[c]{@{}p{6cm}@{}} 
Select samples that maximize the minimum distance to the seed data in feature, output, latent, or an embedding space to avoid redundancy and maximize diversity.
\end{tabular}

& \begin{tabular}[c]{@{}p{6cm}@{}} 
Discovery of high-entropy oxides for \ce{H2} production \cite{nie2024active}; 
Surface structure exploration for catalysis \cite{jung2023machine}
\end{tabular} \\

& Physical metric-based
& \begin{tabular}[c]{@{}p{6cm}@{}} 
Select batches of samples that are mutually dissimilar while also distinct from the existing labeled data based on a physical or chemically motivated metric
\end{tabular}

& \begin{tabular}[c]{@{}p{6cm}@{}} 
Developing accurate property-prediction models for structure-property mapping of microstructures \cite{liu2024active}
\end{tabular} \\
\hline

\end{tabular}
\end{adjustbox}
\end{table*}

\textbf{Model-free vs. model-based active learning :} While most AL workflows employed in materials science are built around surrogate ML models and consequently face the challenges outlined above, model-free strategies could also be adapted for conducting AL as an alternative. A key advantage of model-free strategies is their conceptual simplicity and flexibility; sampling decisions can be designed to target specific objectives without relying on model predictions or uncertainty estimates. This is particularly attractive for scenarios involving a very small amount of seed data, where surrogate models have limited predictive accuracy, and their uncertainty estimates may be unreliable. For example, Information-theory based metrics such as information-entropy and mutual information capture expected information gain beyond model-derived properties \cite{sourati2016classification,wu2023fusing}. Zhang et al. \cite{zhang2023entropy} employed an information-entropy-based AL workflow to mitigate structure–stability bias in computational crystal databases, where low-symmetry structures are often underrepresented. By prioritizing structurally informative samples, measured using information entropy, their approach improved the coverage in crystallographic space and yielded ML models with superior predictive performance compared to random sampling (Fig. 2c). Similarly, Schwalbe-Koda et al. \cite{schwalbe2025model} demonstrated that atomistic information entropy, computed directly from local atomic descriptors, can serve as a model-free proxy for uncertainty of MLIPs, guiding MD simulations. Beyond informativeness, model-free AL can explicitly enhance representativeness and diversity, two criteria that are often weakly controlled in model-based AL workflows. Density-based strategies, such as clustering or kernel density estimation, promote sampling from statistically significant regions of the materials space, thereby preserving global coverage and facilitating the representative sampling \cite{gkatsis2025density,kim2022defense}.  On the other hand, similary-based model-free strategies emphasize improving  diversity and minimizing redundancy. These methods are often implemented using some distance-based metrics, defined over feature, descriptor, latent, or embedding spaces \cite{janet2019quantitative,li2023exploiting,li2024md} and select samples that are maximally distinct from one another and from the existing seed data, promoting diversity. Alternatively, diversity can also be enforced using physically or chemically motivated similarity measures, for example, based on composition, local coordination environments, structural topology, bonding motifs, or symmetry classes \cite{li2024md,liu2024active}.
It has been shown that when labeled data are scarce, similarity-based model-free methods can outperform model-based AL due to their robustness against inaccurate surrogate models \cite{zhan2021comparative}. However, these approaches are not without limitations. High-dimensional vector spaces  used to represent materials data may suffer from the \textit{curse of dimensionality}, and outcomes of distance-based sampling are sensitive to the choice of representations (e.g. elemental-property-based features or SOAP descriptors \cite{bartok2013representing}), similarity metrics (e.g. Euclidean or Mahalanobis distances), and analytical choices such as centroid-based versus nearest-neighbor selection \cite{omee2024structure}. Recent benchmarking by Bi et al. \cite{bi2025comprehensive} further suggests that, on average, model-free strategies underperform model-based AL when evaluated across diverse materials datasets, as they lack explicit mechanisms to capture the relationship between samples and target properties. Despite these limitations, model-free AL remains practically valuable; it avoids the computational overhead of training and retraining surrogate models (e.g. ensembles of ML models) and remains effective when initial datasets are small or highly biased. These observations highlight an unresolved dilemma in AL design: whether to prioritize model-based or model-free strategies, and motivate hybrid approaches that combine the robustness of model-free sampling with the task-awareness of model-based methods \cite{jose2024regression,kee2018query}. Representative model-free AL strategies and their applications in materials science are summarized in Table \ref{Table_ModelFreeAL}.

\begin{table*}[!ht]
\centering
\caption{Common surrogate models and their properties relevant to Bayesian Optimization based active learning applications. The \textit{exp} and \textit{comp} in parenthesis of literature references indicates whether the works do involve experiments or purely computational simulations, respectively.}
\captionsetup{justification=raggedright, singlelinecheck=false}
\begin{adjustbox}{width=1\textwidth}
\renewcommand{\arraystretch}{1.5}
\begin{tabular}{lccccc}
\hline
\textbf{Model Type} & \textbf{Data Efficiency} & \textbf{UQ} & \textbf{Interpretability} & \textbf{Cost} & \textbf{Application in Materials Science}\\
\hline
Gaussian Processes & High & Principled (exact Bayesian posterior) & Limited in high-dimension & High ($\mathcal{O}(N^3)$) &  \begin{tabular}[c]{@{}p{5cm}@{}}Phase-change memory material for photonic switching devices \cite{kusne2020fly} (exp); layered materials with suitable electronic properties \cite{bassman2018active} (comp)\end{tabular} \\
Random Forests & Moderate &  Heuristic (ensemble-based) & Moderate with feature importance & Low & \begin{tabular}[c]{@{}p{5cm}@{}} Biochar syntehsis for \ce{CO2} capture \cite{yuan2024active} (exp) ; screening of inorganic materials  \cite{min2020accelerated} (comp) \end{tabular}
\\
Gradient Boosting Methods & Moderate & Heuristic (ensemble-based) & Moderate with feature importance & Moderate & \begin{tabular}[c]{@{}p{5cm}@{}} High-entropy oxides for \ce{H2} production  \cite{nie2024active} (exp) ; power factor prediction of thermoelectrics \cite{sheng2020active} (comp) 
 \end{tabular}
\\
Bayesian Neural Networks & Low & Heuristic (approximate posterior) & Low & High & \begin{tabular}[c]{@{}p{5cm}@{}} Optimal parameters for chemical reactions \cite{hase2018phoenics} (exp); van der
Waals heterostructures with suitable bandgaps  \cite{fronzi2021active} (comp) \end{tabular} 
\\
Support Vector Regression & Moderate & Limited & Limited & Moderate & \begin{tabular}[c]{@{}p{5cm}@{}} Shape memory alloys with low thermal hysteresis \cite{xue2016accelerated} (exp) ; Piezoelectric materials screening  \cite{tian2020role} (comp) \end{tabular}
\\
Deep ensembles & Low & Heuristic (inter-model predictive variance) & Low & High & \begin{tabular}[c]{@{}p{5cm}@{}} Crystal structure prediction \cite{hessmann2025accelerating}  (comp); MLIP assisted material simulations \cite{zhang2019active} (comp) \end{tabular} 
\\ 
Symbolic regression & High & Limited & High (analytical equations) & High & \begin{tabular}[c]{@{}p{5cm}@{}} Screening of acid-stable oxides for electrocatalysis \cite{nair2025materials} (comp) \end{tabular} 
\\
\hline
\end{tabular}
\end{adjustbox}
\label{Table_surrogates}
\end{table*}

\vspace{-15pt}
\subsection{Materials Property Optimization and Discovery}
\textbf{Interplay between surrogate models and sampling strategy}: The dominant AL practices in both computational and experimental settings for materials property optimization and discovery are black-box optimization (BBO), with Bayesian Optimization (BO) \cite{shahriari2015taking} being the most widely adopted approach \cite{wu2024race}. BO’s ability to navigate complex search spaces under data-scarce conditions has led to numerous success stories in materials discovery \cite{deshwal2021bayesian, honarmandi2022accelerated, pedersen2021bayesian}. Notable examples include the identification of Pb-free \ce{BaTiO3}-based piezoelectrics with enhanced electrostrictive strain \cite{yuan2018accelerated}, NiTi-based shape memory alloys with low thermal hysteresis \cite{xue2016accelerated}, and efficient high-entropy alloy catalysts \cite{pedersen2021bayesian}. The key components of a BO-driven AL (BO-AL) include: (i) a surrogate model to approximate the expensive objective function mapping the materials property to a set of given input parameters (ii) an acquisition function (analogous to sampling strategy in non-BO AL) to guide sample selection, and (iii) an oracle for labelling new data points. While BO traditionally relies on probabilistic surrogate models such as Gaussian Processes (GPs) for their principled uncertainty quantification, non-Bayesian models have also been adopted in materials applications \cite{lim2021extrapolative, liang2021benchmarking}. This is often motivated by practical considerations, including the poor scalability of GPs in high-dimensional spaces and the superior extrapolation performance observed with certain non-Bayesian models on specific datasets \cite{moriconi2020high}. Nevertheless, the optimal choice of surrogate model for BO-AL in materials science remains an open question, with conflicting findings reported in the literature. For instance, Lim et al. \cite{lim2021extrapolative} demonstrated that GPs with carefully selected kernels outperformed alternative models, including random forests (RF), on experimental materials datasets. In contrast, Liang et al. \cite{liang2021benchmarking} found that RF-based BO outperformed GP-BO using standard isotropic kernels and performed comparably to GPs with anisotropic kernels. Table \ref{Table_surrogates} summarizes some of the characteristics of various surrogate models used in BO-AL for materials science applications.
\\

Importantly, the performance of BO-AL is often not governed just by the surrogate model or acquisition function in isolation, but by their combined behaviour. A highly accurate surrogate model may still perform poorly if paired with a suboptimal sampling strategy. For example, expected improvement (EI), a popular acquisition function used in BO-AL for material property optimization, depends on the current best observation, which might be an unreliable benchmark if the seed data is strongly biased,  misleading the optimization trajectory. Therefore, evaluating surrogate model–acquisition function combinations through after-the-fact (AFT) AL trials (where pool data is already labelled but excluded from seed data) can help identify optimal configurations \cite{xue2016accelerated}, though these analyses may not always generalize to all pool datasets due to distribution shifts. As illustrated by Boley et al. \cite{bauer2024roadmap} (Fig. 3a), in the AFT-AL experiment for discovering perovskites with high bulk modulus, RF with both pure exploitation (XT) and EI acquisition functions perform similarly to that of GP with EI, but only as good as RS. However, in real AL runs (using DFT calculations as ground truth), GP with EI clearly outperforms the others, highlighting that surrogate model–acquisition function selection based solely on initial data may be misleading due to distributional shifts and the underrepresentation of high-performing materials. Although dynamic switching between acquisition functions has been proposed \cite{benjamins2022pi}, it remains underexplored in BO-AL workflows applied for materials discovery. It is to be noted that such a sensitivity to sampling strategy is a general challenge in AL, even beyond BO frameworks. Liu et al. \cite{liu2024active} demonstrated this by analyzing how representation and sampling methods affect the selection of microstructures for property prediction via AL. They found that a sampling strategy considering both input features and model outputs outperformed those based on either alone, including uncertainty-based approaches, highlighting the crucial role of the sampling strategy in determining the AL efficiency.

\begin{figure*}[ht]
    \centering
    \includegraphics[width=\linewidth]{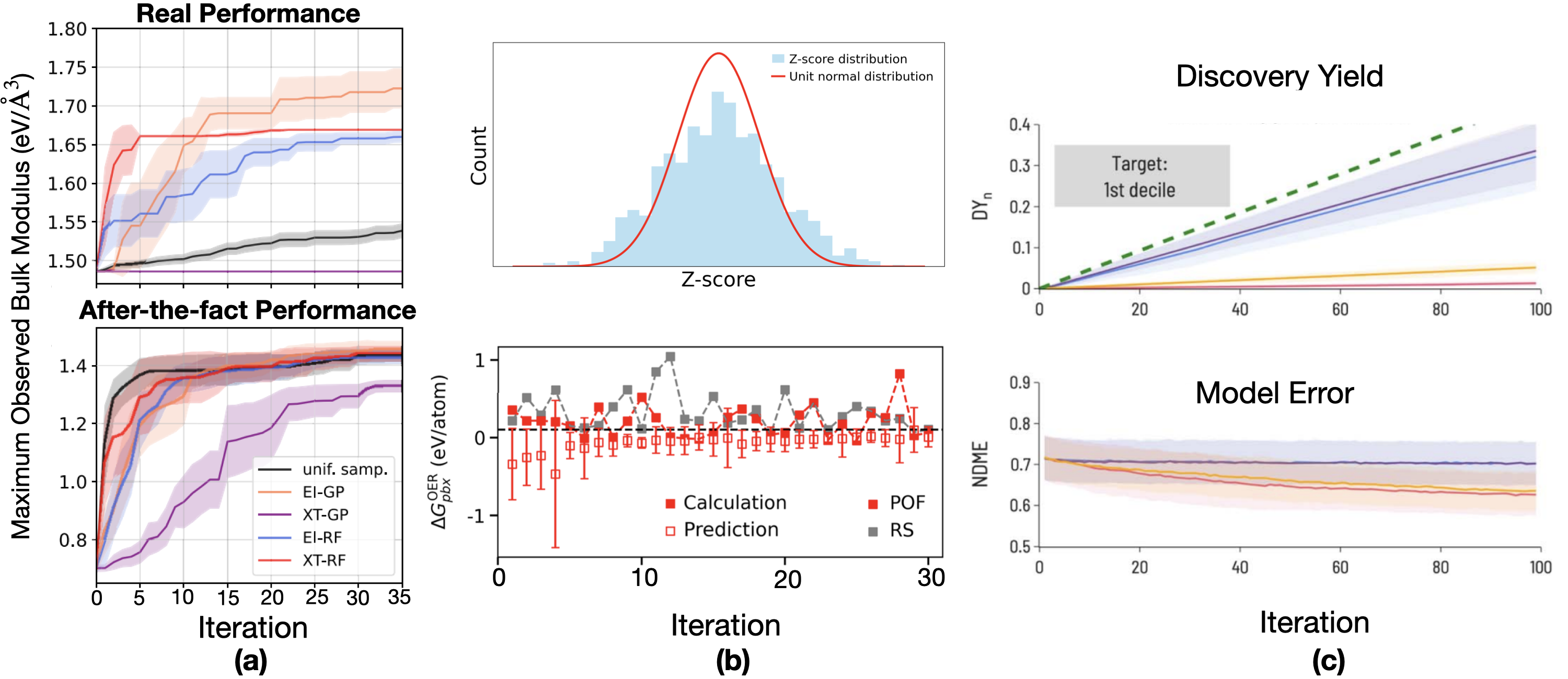}
    \captionsetup{justification=justified}
Fig. 3: (a) Impact of surrogate model and acquisition function choices on active learning for identifying perovskites with high bulk modulus: Real AL runs (top panel) show varied efficiencies across GP and RF models with exploitation (XT), expected improvement (EI), and uniform (random) sampling acquisition functions, compared to retrospective seed-only baselines (bottom panel). Adapted with permission from Ref. \cite{bauer2024roadmap}, Copyright 2024 Institute of Physics.
(b) Unreliable uncertainty estimates: deviation of z-score distributions (residuals scaled by predicted uncertainty) from the unit normal indicates overconfidence (top panel). In AL-guided computational search of acid-stable oxides with the promising figure of merit (Pourbaix decomposition free energy $\Delta G_{pbx} \leq 0.1$ eV/atom), such overconfident estimates are frequent, despite the success of AL in identifying 12 acid-stable oxides within 30 iterations guided by SISSO \cite{ouyang2018sisso} as surrogate model and probability of feasibility (POF) as acquisition function (bottom panel). Reprinted with permission from Ref. \cite{nair2025materials}, Copyright 2025 Springer, (c) Lack of correlation between model performance and materials discovery in Bayesian optimization based AL for identifying first decile materials with near-zero band gaps in the dataset. Colored curves denote acquisition functions (blue: expected value, violet: EI, red: maximum uncertainty, yellow: random sampling). Discovery yield (DY) (see Table III for definition) improves, but non-dimensional model error (NDME = RMSE/standard deviation of holdout set) does not consistently decrease. Adapted with permission from Ref. \cite{borg2023quantifying}, Copyright 2023 Royal Society of Chemistry
\label{Fig:3}
\end{figure*}

For many applications, BO-AL is preferred to be applied in batched or parallel settings where a number of samples (instead of one as in sequential BO) are selected, which could leverage the availability of oracles with parallel execution capabilities (e.g. high performance computing facilities, high-throughput synthesis platforms)  \cite{ozbayram2025batch,hase2018phoenics}. While batching is desirable from a resource management perspective, it can risk degrading sample diversity and informativeness unless the sampling strategy is designed to account for these factors \cite{fromer2025batched}. Acquisition functions often fail to correctly rank batched instances in terms of marginal utility, potentially introducing redundancy or bias.  This is particularly problematic in materials science, where multiscale material behaviours and divergent synthesis/characterization routes can make it difficult to evaluate a batch under uniform experimental conditions. Nevertheless, platforms such as Pheonix \cite{hase2018phoenics}, ChemOS \cite{roch2018chemos}, Olympous \cite{hase2021olympus} etc. have made success in enabling the application of BO-AL with robotic experimentation or lab automation systems to enable parallel acquisition and execution, useful for chemistry and materials science. 

\textbf{Unreliable uncertainty quantification}: In AL strategies with an exploration component such as those used in BO, uncertainty quantification (UQ) plays a pivotal role. This is based on the notion that the surrogate models are most error-prone in regions where their predictions are least confident, and hence, acquiring data from such regions is often the most informative. Reliable UQ helps in identifying these regions by assigning a confidence interval to predictions, thus guiding the sampling strategy toward high-impact data points. However, two core challenges hinder this process: (i) most ML models lack inherent UQ capabilities (see Table \ref{Table_surrogates}), and (ii) many UQ techniques produce unreliable estimates which are often underconfident or overconfident relative to actual prediction errors (Fig. 3b) \cite{pernot2023calibration, hwang2024overcoming}. Unreliable UQ can misguide AL by undervaluing informative samples, potentially leading to premature convergence or excessive exploitation of suboptimal regions. For instance, in BO-AL with EI as the acquisition function, overconfident uncertainty estimates shrink the exploration term, causing the algorithm to overlook uncertain but informative regions in favor of already well-explored ones.

\begin{table*}[!ht]
\centering
\caption{Evaluation metrics used for assessing AL performance for materials property optimization and discovery. $\mathrm{i^{AL}}$ is the active leaning iteration(s) and $\mathrm{i^{RS}}$ is the random sampling iterations. The figure of merit (FOM) is a general quantitative metric measuring the AL efficiency and could indicate the desired material property value, the number of promising materials in a target range, etc. The previous studies employed the metrics for performance evaluation of AL are indicated.}
\captionsetup{justification=raggedright, singlelinecheck=false}
\label{Table_metrics}
\begin{adjustbox}{max width=\textwidth}
\renewcommand{\arraystretch}{2.0}
\begin{tabular}{lcc}
\hline
\textbf{Metric} & \textbf{Definition} & \textbf{Limitation}
\\
\hline
Discovery yield \cite{rohr2020benchmarking}, \cite{borg2023quantifying}  &
$\frac{\text{No. of Materials with FOM at } i^{\mathrm{AL}}}{\text{Total No. of Materials with FOM in Pool}}$ &
\begin{tabular}[c]{@{}p{5cm}@{}} 
Requires apriori knowledge of
interesting materials in the pool
\end{tabular}
\\
Acceleration factor \cite{liang2021benchmarking},\cite{borg2023quantifying} & 
$\frac{\text{No. of } i^{\mathrm{AL}} \text{ for FOM}}{\text{No. of } i^{\mathrm{RS}} \text{ for FOM}}$

& Not (always) a fair baseline 
\\
Enhancement factor \cite{liang2021benchmarking},\cite{borg2023quantifying} &  $\frac{\mathrm{FOM(AL)}}{\mathrm{FOM(RS)}}$ & Not (always) a fair baseline 
\\

Decision eﬃciency \cite{rohr2020benchmarking} & 
\begin{tabular}[c]{@{}l@{}}
$2 \cdot \frac{\#\{ \text{samples with } \mathrm{i_{FOM}} \leq \mathrm{i_{FOM}^{(selected)}} \}}{N} -1 $ 
\end{tabular} &
Dependence on pool quality 
\\ 
Model performance \cite{li2023exploiting}, \cite{kavalsky2023much} & 
\begin{tabular}[c]{@{}p{5cm}@{}} 
\quad Error of surrogate model \\[-2ex]
estimated on a (holdout) test set
\end{tabular}  &  
\begin{tabular}[c]{@{}p{5cm}@{}} 
\quad\quad Prone to overestimation \\[-2ex]
\quad\quad due to active learning bias
\end{tabular}
\\

Model uncertainty \cite{nair2025materials}, \cite{kavalsky2024multiobjective}  &
Uncertainty of surrogate model  &
\begin{tabular}[c]{@{}p{5cm}@{}}
\quad\quad Prone to misleading \\[-2ex]
\quad\quad due to unreliable uncertainties 
\end{tabular}
\\
\hline
\end{tabular}
\end{adjustbox}
\label{Table_metrics}
\end{table*}

The reliability of UQ in ML remains an open research challenge in materials science, with relatively few studies that critically benchmark UQ methods \cite{varivoda2023materials, tran2020methods}. Existing efforts typically evaluate the quality of uncertainty estimates based on hold-out test sets using metrics based on the uncertainty distribution \cite{gruich2023clarifying}, correlation with prediction errors \cite{varivoda2023materials}, or computational cost \cite{tran2020methods}. These studies consistently show that no single UQ method outperforms across all scenarios, with strong dependencies on surrogate model and dataset characteristics. Nevertheless, calibration approaches such as scaling uncertainty estimates with respect to residuals can enhance their reliability \cite{palmer2022calibration, gruich2023clarifying}. Openly available tools like uncertainty-toolbox \cite{chung2021uncertainty} and UQLab \cite{lataniotis2022uqlab} facilitate such calibration processes. However, the effectiveness of these calibration methods within AL workflows remains largely unexplored, with existing studies still in their early stages \cite{thomas2023calibration}. Since such approaches require labeled data, they are constrained to the currently known samples  and hence cannot guarantee improved uncertainty estimates on the unlabeled pool data where accurate UQ is most critical. This highlights the need for more standardized, AL-focused metrics to assess and compare UQ methods for materials property prediction and optimization tasks.

Since one of the reasons for unreliable UQ is that the surrogate models are primarily trained to minimize prediction error without providing default uncertinay estimates, a range of  alternative, model-agnostic UQ strategies have been proposed to address this challenge. These rely on heuristic measures of uncertainty, such as distance in feature \cite{korolev2022universal} or latent spaces \cite{janet2019quantitative} and are not specifically dependent on predictive model architecture. It has been demonstrated that latent space distance provides more reliable uncertainty estimates (by better capturing of residuals) for both artificial \cite{janet2019quantitative} and graph \cite{musielewicz2024improved} neural networks, indicating that such model-agnostic methods could be beneficial to AL. Similarly, integrating Domain of Applicability (DoA) analysis \cite{schultz2025general, hu2024realistic} with UQ in AL workflows allows for the identification and expansion of the model’s reliable prediction sregime, enabling more effective sample selection near or beyond the current DoA to improve model robustness \cite{zhong2022enlarging}. However, these methods are more mature in cheminformatics, and lack standardized implementation in materials science, due to the absence of universally accepted DoA frameworks \cite{sutton2020identifying,schultz2025general}. Also, a fair comparison between these model-agnostic  vs. model-based UQ methods for AL performance for materials property optimization has not been done to the best of our knowledge, which is also necessary for improving the reliability of AL practices in materials science.

\textbf{Lack of standardized performance evaluation metrics}: A rigorous and accurate evaluation of AL performance is essential for assessing its efficiency and determining when to terminate the AL loop. This is especially critical in closed-loop, automated experimentation platforms with limited human intervention, where failure to recognize diminishing returns can result in substantial resource waste. Various performance metrics have been proposed to quantify AL efficiency in materials property optimization. However, no single metric universally captures all aspects of performance, and even metrics which are expected to be complementary could show diverging trends for performance estimation throughout AL iterations. The most popular metrics used in assessing the efficacy of AL are based on comparison with random sampling (RS) \cite{lim2021extrapolative,kavalsky2023much}, which involves sampling data points from the pool without the use of a surrogate model derived insights and hence also can be referred to as \qq{passive learning}. Rohr et al. \cite{rohr2020benchmarking} proposed quantitative metrics such as the acceleration and enhancement factors (see Table III), measuring the fraction of promising materials found and the iterations required to find them relative to RS, which have been used to benchmark surrogate models and acquisition strategies for AFT materials discovery campaigns \cite{liang2021benchmarking, palizhati2022agents}. However, using RS as a baseline may not be considered a universal performance metric in materials science, where experimental choices are typically guided by prior knowledge rather than random selection. Moreover, although not widely reported in materials property optimization, AL has been shown to occasionally underperform RS in other domains \cite{schein2007active, fuhg2021state}, which could arise due to inadequate surrogate model and/or acquisition function choices.

Another common evaluation criterion in materials property optimization and AL more broadly is tracking model performance over AL iterations. However, Borg et al. \cite{borg2023quantifying} observed that improvements in model accuracy do not necessarily correlate with better discovery rates of high-performing materials; in some cases, less accurate models identified promising candidates more effectively (Fig. 3c). This suggests that the popular notion of \qq{model is getting better} may not qualify as an accurate evaluation criteria for AL for materials discovery. Similarly, Koizumi et al.\cite{koizumi2024performance} demonstrated that the AL performance can vary substantially with changes in the material property, descriptor dimension and size of seed data even when using a fixed choice for surrogate model and sampling strategy. Kim et al. \cite{kim2020machine} emphasized that beyond the model and seed data, the quality of the candidate pool, particularly the fraction of promising materials, critically affects AL performance evaluation. They pointed out that materials property optimization depend on how likely candidates in the pool are to outperform those in the initial seed data and proposed metrics such as \textit{predicted fraction of improved candidates} to quantify the pool quality. It is to be noted that all of these insights stem from post hoc  studies using fully labeled datasets and are not based on AL workflows involving real-time experiments or simulations. Nevertheless, the use of diverse datasets spanning different properties, surrogate models, and acquisition functions, along with repeated trials to gather statistics suggests that these trends may hold more broadly, including for AL workflows involving real-time experiments and simulations. 

\begin{algorithm}[h!]
\caption{Proposed Hierarchical Sampling Scheme for General Active Learning Sampling}\label{algorithm:taskaware}
\begin{algorithmic}[1]
\Require Unlabeled candidate pool $\mathcal{U} = \{x_i\}$, representation $\phi(x)$, 
informativeness scoring function $I(x)$ sample size $S$, 
physical validity filter $F_\text{phys}$, cost function $C(x)$, representativeness threshold $R_\text{min}$

\Ensure Selected samples $\mathcal{S} \subseteq \mathcal{U}$ satisfying multiple criteria

\State \textbf{Step 1 — Physical \& Feasibility Filtering:}  
\State $\mathcal{U}_\text{valid} \gets \{ x \in \mathcal{U} \mid F_\text{phys}(x) = \text{True} \wedge C(x) \le C_\text{max} \}$  
\Comment{Enforces physically meaningful, cost-feasible candidates}

\State \textbf{Step 2 — Informativeness Scoring:}  
\State Compute $I(x)$ for each $x \in \mathcal{U}_\text{valid}$  
\Comment{Model-based or model-free measure of expected information gain}

\State \textbf{Step 3 — Preselection of Top Candidates:}  
\State $\mathcal{U}_\text{top} \gets$ top-$K$ samples ranked by $I(x)$  
\Comment{Limits to informative subset, avoids extreme outliers}

\State \textbf{Step 4 — Representativeness Filtering:}  
\State Compute representativeness score $R(x)$ (e.g. local density or cluster-based weight) for $x \in \mathcal{U}_\text{top}$  
\State $\mathcal{U}_\text{rep} \gets \{ x \in \mathcal{U}_\text{top} \mid R(x) \ge R_\text{min} \}$  
\Comment{Ensures coverage}

\State \textbf{Step 5 — Diversity Selection (Sample Size $S$):}  
\State Select $\mathcal{S} \subseteq \mathcal{U}_\text{rep}$ of size $S$ to maximize pairwise dissimilarity:  
\[
\mathcal{S} = \arg\max_{\substack{\mathcal{S} \subseteq \mathcal{U}_\text{rep} \\ |\mathcal{S}| = S}} \min_{x_i \neq x_j \in \mathcal{S}} d(\phi(x_i), \phi(x_j))
\]  
\Comment{$d(\cdot,\cdot)$ can be distance-based or physical metric}

\State \Return $\mathcal{S}$
\end{algorithmic}
\end{algorithm}

In AL frameworks focused on multi-fidelity or multi-objective optimization, performance metrics must be adapted beyond those used in single-fidelity and single-objective tasks. In the multi-fidelity setting, data acquisition comes with differing costs and fidelities, making it essential to quantify not just accuracy, but also cost-efficiency. Jacobs et al.\cite{jacobs2023role} demonstrated that low-fidelity data can improve optimization only when it is significantly cheaper (e.g. $\leq$5\% of high-fidelity cost) and partially available upfront. This underscores the need for performance metrics that account for both data cost and availability. While initial efforts exist in this direction \cite{sabanza2024best, magar2023learning}, general-purpose cost functions and robust sampling strategies for multi-fidelity materials optimization remain underdeveloped. In multi-objective AL, the aim is to identify candidate materials that perform well across multiple properties. Most studies have centered on Pareto optimization, where the goal is to approximate the Pareto front of non-dominated solutions \cite{trinquet2025optical, xu2025multi}. However, this approach is most effective when objectives are conflicting, which may not always be the case in materials science. For non-conflicting or correlated properties, scalarization (e.g. using weighted sums) may offer better performance, yet it introduces subjectivity in weight selection and risks missing diverse solutions. Furthermore, materials properties are often non-convex, unevenly distributed, and may vary in measurement cost, complicating both the optimization and its evaluation. These challenges highlight the pressing need for task-aware and cost-sensitive AL performance metrics that go beyond standard error or accuracy. Future efforts should aim to develop generalizable evaluation frameworks that reflect the realities of materials design—such as cost disparities, fidelity trade-offs, and property-specific constraints.

\section{Outlook} 
Active learning has shown substantial promise for accelerating materials simulation and discovery, yet its broader impact is still limited by unresolved challenges across the design and deployment of AL workflows. Overcoming these limitations requires development of better practices for the design and choice of various components of the AL pipeline. A key opportunity lies in the development of AL workflows that incorporate domain knowledge and practical constraints. For example, rather than focusing on a single sampling criterion, future strategies should increasingly adopt hierarchical schemes that jointly account for informativeness, representativeness, diversity, etc. an example for which is demonstrated in Algorithm 2. Such considerations are especially critical in materials science, where unphysical candidates or prohibitively expensive evaluations can quickly undermine the benefits of AL. Equally important is the need for clearly defining the set of hyperparametres (e.g. surrogate model, acqusition function choices) specific to AL and developing software tools that allow for their optimization. In addition, establishing standardized AL pipelines and conducting large-scale benchmarking across diverse materials datasets will be essential for deriving transferable guidelines and reducing reliance on ad hoc heuristics. Looking ahead, several emerging directions offer promising pathways to address persistent bottlenecks. The unreliability of surrogate models in early AL stages may be mitigated by leveraging pretrained or foundation models \cite{batatia2025foundation} trained on related materials properties, enabling more informed sampling with limited data. Reformulating AL as a sequential decision-making problem, such as through reinforcement learning, may further enable adaptive sampling policies in high-dimensional design spaces where conventional approaches struggle \cite{xian2025unlocking}. In parallel, recent advances in generative models open opportunities to couple AL with candidate generation, allowing efficient exploration of vast design spaces while iteratively steering generation toward high-performing regions \cite{xin2021active,schwarting2025steering}. The progress in the design of AL workflows also needs a stronger integration between the materials and ML or communities as many advances relevant to AL such as uncertainty quantification developed by the later remain underexplored for AL for materials science applications. Bridging this gap through interdisciplinary efforts will be crucial for translating methodological advances into robust, scalable AL workflows capable of addressing the complexity and cost constraints of real-world materials applications.




\section*{Data Availability}
No data is newly generated during this work. 

\section*{Acknowledgement}
A.S.N thanks Matthias Scheffler for helpful discussions, and Beate Paulus and the German Research Foundation (DFG) for support through the Walter-Benjamin Fellowship Program (project No. 540316537).

\section*{References}
\bibliography{reference.bib}

@inproceedings{campbell2000query,
  title={Query learning with large margin classifiers},
  author={Campbell, Colin and Cristianini, Nello and Smola, Alex and others},
  booktitle={ICML},
  volume={20},
  number={0},
  pages={0},
  year={2000}
}

@article{hwang1991query,
  title={Query-based learning applied to partially trained multilayer perceptrons},
  author={Hwang, Jenq-Neng and Choi, Jai J and Oh, Seho and Marks, RJ and others},
  journal={IEEE Transactions on Neural Networks},
  volume={2},
  number={1},
  pages={131--136},
  year={1991}
}

@article{khosravi2024data,
  title={A data driven sequential learning framework to accelerate and optimize multi-objective manufacturing decisions},
  author={Khosravi, Hamed and Olajire, Taofeeq and Raihan, Ahmed Shoyeb and Ahmed, Imtiaz},
  journal={Journal of Intelligent Manufacturing},
  pages={1--26},
  year={2024},
  publisher={Springer}
}

@article{rohr2020benchmarking,
  title={Benchmarking the acceleration of materials discovery by sequential learning},
  author={Rohr, Brian and Stein, Helge S and Guevarra, Dan and Wang, Yu and Haber, Joel A and Aykol, Muratahan and Suram, Santosh K and Gregoire, John M},
  journal={Chemical science},
  volume={11},
  number={10},
  pages={2696--2706},
  year={2020},
  publisher={Royal Society of Chemistry}
}

@article{jain2013commentary,
  title={Commentary: The Materials Project: A materials genome approach to accelerating materials innovation},
  author={Jain, Anubhav and Ong, Shyue Ping and Hautier, Geoffroy and Chen, Wei and Richards, William Davidson and Dacek, Stephen and Cholia, Shreyas and Gunter, Dan and Skinner, David and Ceder, Gerbrand and others},
  journal={APL materials},
  volume={1},
  number={1},
  year={2013},
  publisher={AIP Publishing}
}

@article{wang2022benchmarking,
  title={Benchmarking active learning strategies for materials optimization and discovery},
  author={Wang, Alex and Liang, Haotong and McDannald, Austin and Takeuchi, Ichiro and Kusne, Aaron Gilad},
  journal={Oxford Open Materials Science},
  volume={2},
  number={1},
  pages={itac006},
  year={2022},
  publisher={Oxford University Press}
}

@article{kulichenko2024data,
  title={Data generation for machine learning interatomic potentials and beyond},
  author={Kulichenko, Maksim and Nebgen, Benjamin and Lubbers, Nicholas and Smith, Justin S and Barros, Kipton and Allen, Alice EA and Habib, Adela and Shinkle, Emily and Fedik, Nikita and Li, Ying Wai and others},
  journal={Chemical Reviews},
  volume={124},
  number={24},
  pages={13681--13714},
  year={2024},
  publisher={ACS Publications}
}

@article{bhatia2025leveraging,
  title={Leveraging active learning-enhanced machine-learned interatomic potential for efficient infrared spectra prediction},
  author={Bhatia, Nitik and Rinke, Patrick and Krej{\v{c}}{\'\i}, Ond{\v{r}}ej},
  journal={npj Computational Materials},
  volume={11},
  number={1},
  pages={324},
  year={2025},
  publisher={Nature Publishing Group UK London}
}

@article{wang2015querying,
  title={Querying discriminative and representative samples for batch mode active learning},
  author={Wang, Zheng and Ye, Jieping},
  journal={ACM Transactions on Knowledge Discovery from Data (TKDD)},
  volume={9},
  number={3},
  pages={1--23},
  year={2015},
  publisher={ACM New York, NY, USA}
}

@article{du2015exploring,
  title={Exploring representativeness and informativeness for active learning},
  author={Du, Bo and Wang, Zengmao and Zhang, Lefei and Zhang, Liangpei and Liu, Wei and Shen, Jialie and Tao, Dacheng},
  journal={IEEE transactions on cybernetics},
  volume={47},
  number={1},
  pages={14--26},
  year={2015},
  publisher={IEEE}
}

@article{huang2010active,
  title={Active learning by querying informative and representative examples},
  author={Huang, Sheng-Jun and Jin, Rong and Zhou, Zhi-Hua},
  journal={Advances in neural information processing systems},
  volume={23},
  year={2010}
}

@inproceedings{li2012active,
  title={Active learning for hierarchical text classification},
  author={Li, Xiao and Kuang, Da and Ling, Charles X},
  booktitle={Pacific-Asia conference on knowledge discovery and data mining},
  pages={14--25},
  year={2012},
  organization={Springer}
}

@article{zaverkin2022exploring,
  title={Exploring chemical and conformational spaces by batch mode deep active learning},
  author={Zaverkin, Viktor and Holzm{\"u}ller, David and Steinwart, Ingo and K{\"a}stner, Johannes},
  journal={Digital Discovery},
  volume={1},
  number={5},
  pages={605--620},
  year={2022},
  publisher={Royal Society of Chemistry}
}

@inproceedings{brinker2003incorporating,
  title={Incorporating diversity in active learning with support vector machines},
  author={Brinker, Klaus},
  booktitle={Proceedings of the 20th international conference on machine learning (ICML-03)},
  pages={59--66},
  year={2003}
}

@article{sourati2016classification,
  title={Classification active learning based on mutual information},
  author={Sourati, Jamshid and Akcakaya, Murat and Dy, Jennifer G and Leen, Todd K and Erdogmus, Deniz},
  journal={Entropy},
  volume={18},
  number={2},
  pages={51},
  year={2016},
  publisher={MDPI}
}

@article{gruich2023clarifying,
  title={Clarifying trust of materials property predictions using neural networks with distribution-specific uncertainty quantification},
  author={Gruich, Cameron J and Madhavan, Varun and Wang, Yixin and Goldsmith, Bryan R},
  journal={Machine Learning: Science and Technology},
  volume={4},
  number={2},
  pages={025019},
  year={2023},
  publisher={IOP Publishing}
}

@article{tran2020methods,
  title={Methods for comparing uncertainty quantifications for material property predictions},
  author={Tran, Kevin and Neiswanger, Willie and Yoon, Junwoong and Zhang, Qingyang and Xing, Eric and Ulissi, Zachary W},
  journal={Machine Learning: Science and Technology},
  volume={1},
  number={2},
  pages={025006},
  year={2020},
  publisher={IOP Publishing}
}

@article{kee2018query,
  title={Query-by-committee improvement with diversity and density in batch active learning},
  author={Kee, Seho and Del Castillo, Enrique and Runger, George},
  journal={Information Sciences},
  volume={454},
  pages={401--418},
  year={2018},
  publisher={Elsevier}
}

@article{wu2023fusing,
  title={Fusing information entropy and similarity: A novel active learning strategy for chemical process fault classifications},
  author={Wu, Shuhui and Zhao, Zihao and Yin, Min and Li, Hongguang},
  journal={Chemometrics and Intelligent Laboratory Systems},
  volume={237},
  pages={104821},
  year={2023},
  publisher={Elsevier}
}

@article{xue2016accelerated,
  title={Accelerated search for materials with targeted properties by adaptive design},
  author={Xue, Dezhen and Balachandran, Prasanna V and Hogden, John and Theiler, James and Xue, Deqing and Lookman, Turab},
  journal={Nature communications},
  volume={7},
  number={1},
  pages={1--9},
  year={2016},
  publisher={Nature Publishing Group}
}

@article{benjamins2022pi,
  title={Pi is back! Switching acquisition functions in Bayesian optimization},
  author={Benjamins, Carolin and Raponi, Elena and Jankovic, Anja and van der Blom, Koen and Santoni, Maria Laura and Lindauer, Marius and Doerr, Carola},
  journal={arXiv preprint arXiv:2211.01455},
  year={2022}
}

@inproceedings{zhan2021comparative,
  title={A Comparative Survey: Benchmarking for Pool-based Active Learning.},
  author={Zhan, Xueying and Liu, Huan and Li, Qing and Chan, Antoni B},
  booktitle={IJCAI},
  pages={4679--4686},
  year={2021}
}

@article{ozbayram2025batch,
  title={Batch active learning for microstructure--property relations in energetic materials},
  author={Ozbayram, Ozge and Olsen, Daniel and Annamaraju, Maruthi and Robertson, Andreas E and Venkatraman, Aditya and Kalidindi, Surya R and Zhou, Min and Graham-Brady, Lori},
  journal={Mechanics of Materials},
  volume={205},
  pages={105308},
  year={2025},
  publisher={Elsevier}
}

@article{liang2021benchmarking,
  title={Benchmarking the performance of Bayesian optimization across multiple experimental materials science domains},
  author={Liang, Qiaohao and Gongora, Aldair E and Ren, Zekun and Tiihonen, Armi and Liu, Zhe and Sun, Shijing and Deneault, James R and Bash, Daniil and Mekki-Berrada, Flore and Khan, Saif A and others},
  journal={npj Computational Materials},
  volume={7},
  number={1},
  pages={188},
  year={2021},
  publisher={Nature Publishing Group UK London}
}

@article{koizumi2024performance,
  title={Performance of uncertainty-based active learning for efficient approximation of black-box functions in materials science},
  author={Koizumi, Ai and Deffrennes, Guillaume and Terayama, Kei and Tamura, Ryo},
  journal={Scientific Reports},
  volume={14},
  number={1},
  pages={27019},
  year={2024},
  publisher={Nature Publishing Group UK London}
}

@article{magar2023learning,
  title={Learning from mistakes: Sampling strategies to efficiently train machine learning models for material property prediction},
  author={Magar, Rishikesh and Farimani, Amir Barati},
  journal={Computational Materials Science},
  volume={224},
  pages={112167},
  year={2023},
  publisher={Elsevier}
}

@article{li2023exploiting,
  title={Exploiting redundancy in large materials datasets for efficient machine learning with less data},
  author={Li, Kangming and Persaud, Daniel and Choudhary, Kamal and DeCost, Brian and Greenwood, Michael and Hattrick-Simpers, Jason},
  journal={Nature Communications},
  volume={14},
  number={1},
  pages={7283},
  year={2023},
  publisher={Nature Publishing Group UK London}
}

@article{schwalbe2025model,
  title={Model-free estimation of completeness, uncertainties, and outliers in atomistic machine learning using information theory},
  author={Schwalbe-Koda, Daniel and Hamel, Sebastien and Sadigh, Babak and Zhou, Fei and Lordi, Vincenzo},
  journal={Nature Communications},
  volume={16},
  number={1},
  pages={4014},
  year={2025},
  publisher={Nature Publishing Group UK London}
}

@article{pa2025information,
  title={Information-entropy-driven generation of material-agnostic datasets for machine-learning interatomic potentials},
  author={PA Subramanyam, Aparna and Perez, Danny},
  journal={npj Computational Materials},
  volume={11},
  number={1},
  pages={218},
  year={2025},
  publisher={Nature Publishing Group UK London}
}

@article{zhang2023entropy,
  title={ET-AL: entropy-targeted active learning for bias mitigation in materials data},
  author={Zhang, Hengrui and Chen, Wei Wayne and Rondinelli, James M and Chen, Wei},
  journal={Applied Physics Reviews},
  volume={10},
  number={2},
  year={2023},
  publisher={AIP Publishing}
}

@article{butler2018machine,
  title={Machine learning for molecular and materials science},
  author={Butler, Keith T and Davies, Daniel W and Cartwright, Hugh and Isayev, Olexandr and Walsh, Aron},
  journal={Nature},
  volume={559},
  number={7715},
  pages={547--555},
  year={2018},
  publisher={Nature Publishing Group UK London}
}

@article{xu2023small,
  title={Small data machine learning in materials science},
  author={Xu, Pengcheng and Ji, Xiaobo and Li, Minjie and Lu, Wencong},
  journal={npj Computational Materials},
  volume={9},
  number={1},
  pages={42},
  year={2023},
  publisher={Nature Publishing Group UK London}
}

@article{maier2007combinatorial,
  title={Combinatorial and high-throughput materials science},
  author={Maier, Wilhelm F and Stoewe, Klaus and Sieg, Simone},
  journal={Angewandte chemie international edition},
  volume={46},
  number={32},
  pages={6016--6067},
  year={2007},
  publisher={Wiley Online Library}
}

@article{curtarolo2013high,
  title={The high-throughput highway to computational materials design},
  author={Curtarolo, Stefano and Hart, Gus LW and Nardelli, Marco Buongiorno and Mingo, Natalio and Sanvito, Stefano and Levy, Ohad},
  journal={Nature materials},
  volume={12},
  number={3},
  pages={191--201},
  year={2013},
  publisher={Nature Publishing Group UK London}
}

@article{sutton2020identifying,
  title={Identifying domains of applicability of machine learning models for materials science},
  author={Sutton, Christopher and Boley, Mario and Ghiringhelli, Luca M and Rupp, Matthias and Vreeken, Jilles and Scheffler, Matthias},
  journal={Nature communications},
  volume={11},
  number={1},
  pages={4428},
  year={2020},
  publisher={Nature Publishing Group UK London}
}

@article{bauer2024roadmap,
  title={Roadmap on Data-Centric Materials Science},
  author={Bauer, Stefan and Benner, Peter and Bereau, Tristan and Blum, Volker and Boley, Mario and Carbogno, Christian and Catlow, Richard and Dehm, Gerhard and Eibl, Sebastian and Ernstorfer, Ralph and others},
  journal={Modelling and Simulation in Materials Science and Engineering},
  year={2024}
}

@article{cohn1994improving,
  title={Improving generalization with active learning},
  author={Cohn, David and Atlas, Les and Ladner, Richard},
  journal={Machine learning},
  volume={15},
  pages={201--221},
  year={1994},
  publisher={Springer}
}

@article{cohn1996active,
  title={Active learning with statistical models},
  author={Cohn, David A and Ghahramani, Zoubin and Jordan, Michael I},
  journal={Journal of artificial intelligence research},
  volume={4},
  pages={129--145},
  year={1996}
}

@article{lookman2019active,
  title={Active learning in materials science with emphasis on adaptive sampling using uncertainties for targeted design},
  author={Lookman, Turab and Balachandran, Prasanna V and Xue, Dezhen and Yuan, Ruihao},
  journal={npj Computational Materials},
  volume={5},
  number={1},
  pages={21},
  year={2019},
  publisher={Nature Publishing Group UK London}
}

@article{bassman2018active,
  title={Active learning for accelerated design of layered materials},
  author={Bassman Oftelie, Lindsay and Rajak, Pankaj and Kalia, Rajiv K and Nakano, Aiichiro and Sha, Fei and Sun, Jifeng and Singh, David J and Aykol, Muratahan and Huck, Patrick and Persson, Kristin and others},
  journal={npj Computational Materials},
  volume={4},
  number={1},
  pages={74},
  year={2018},
  publisher={Nature Publishing Group UK London}
}

@article{gubaev2018machine,
  title={Machine learning of molecular properties: Locality and active learning},
  author={Gubaev, Konstantin and Podryabinkin, Evgeny V and Shapeev, Alexander V},
  journal={The Journal of chemical physics},
  volume={148},
  number={24},
  year={2018},
  publisher={AIP Publishing}
}

@article{merchant2023scaling,
  title={Scaling deep learning for materials discovery},
  author={Merchant, Amil and Batzner, Simon and Schoenholz, Samuel S and Aykol, Muratahan and Cheon, Gowoon and Cubuk, Ekin Dogus},
  journal={Nature},
  volume={624},
  number={7990},
  pages={80--85},
  year={2023},
  publisher={Nature Publishing Group UK London}
}

@article{lyngby2022data,
  title={Data-driven discovery of 2D materials by deep generative models},
  author={Lyngby, Peder and Thygesen, Kristian Sommer},
  journal={npj Computational Materials},
  volume={8},
  number={1},
  pages={232},
  year={2022},
  publisher={Nature Publishing Group UK London}
}

@article{jinnouchi2020fly,
  title={On-the-fly active learning of interatomic potentials for large-scale atomistic simulations},
  author={Jinnouchi, Ryosuke and Miwa, Kazutoshi and Karsai, Ferenc and Kresse, Georg and Asahi, Ryoji},
  journal={The Journal of Physical Chemistry Letters},
  volume={11},
  number={17},
  pages={6946--6955},
  year={2020},
  publisher={ACS Publications}
}

@article{nie2024active,
  title={Active Learning Guided Discovery of High Entropy Oxides Featuring High H2-production},
  author={Nie, Siyang and Xiang, Yan and Wu, Liang and Lin, Guang and Liu, Qingda and Chu, Shengqi and Wang, Xun},
  journal={Journal of the American Chemical Society},
  volume={146},
  number={43},
  pages={29325--29334},
  year={2024},
  publisher={ACS Publications}
}

@article{sumpter2015bridge,
  title={A bridge for accelerating materials by design},
  author={Sumpter, Bobby G and Vasudevan, Rama K and Potok, Thomas and Kalinin, Sergei V},
  journal={NPJ Computational Materials},
  volume={1},
  number={1},
  pages={1--11},
  year={2015},
  publisher={Nature Publishing Group}
}

@article{karande2022strategic,
  title={A strategic approach to machine learning for material science: how to tackle real-world challenges and avoid pitfalls},
  author={Karande, Piyush and Gallagher, Brian and Han, Thomas Yong-Jin},
  journal={Chemistry of Materials},
  volume={34},
  number={17},
  pages={7650--7665},
  year={2022},
  publisher={ACS Publications}
}

@article{hu2024designing,
  title={Designing unique and high-performance Al alloys via machine learning: Mitigating data bias through active learning},
  author={Hu, Mingwei and Tan, Qiyang and Knibbe, Ruth and Xu, Miao and Liang, Guofang and Zhou, Jianxin and Xu, Jun and Jiang, Bin and Li, Xue and Ramajayam, Mahendra and others},
  journal={Computational Materials Science},
  volume={244},
  pages={113204},
  year={2024},
  publisher={Elsevier}
}

@article{suvarna2024active,
  title={Active learning streamlines development of high performance catalysts for higher alcohol synthesis},
  author={Suvarna, Manu and Zou, Tangsheng and Chong, Sok Ho and Ge, Yuzhen and Mart{\'\i}n, Antonio J and P{\'e}rez-Ram{\'\i}rez, Javier},
  journal={Nature Communications},
  volume={15},
  number={1},
  pages={5844},
  year={2024},
  publisher={Nature Publishing Group UK London}
}

@article{jose2024informative,
  title={Informative training data for efficient property prediction in metal--organic frameworks by active learning},
  author={Jose, Ashna and Devijver, Emilie and Jakse, Noel and Poloni, Roberta},
  journal={Journal of the American Chemical Society},
  volume={146},
  number={9},
  pages={6134--6144},
  year={2024},
  publisher={ACS Publications}
}

@article{kang2024accelerating,
  title={Accelerating the training and improving the reliability of machine-learned interatomic potentials for strongly anharmonic materials through active learning},
  author={Kang, Kisung and Purcell, Thomas AR and Carbogno, Christian and Scheffler, Matthias},
  journal={arXiv preprint arXiv:2409.11808},
  year={2024}
}

@article{verdi2021thermal,
  title={Thermal transport and phase transitions of zirconia by on-the-fly machine-learned interatomic potentials},
  author={Verdi, Carla and Karsai, Ferenc and Liu, Peitao and Jinnouchi, Ryosuke and Kresse, Georg},
  journal={npj Computational Materials},
  volume={7},
  number={1},
  pages={156},
  year={2021},
  publisher={Nature Publishing Group UK London}
}

@article{deshwal2021bayesian,
  title={Bayesian optimization of nanoporous materials},
  author={Deshwal, Aryan and Simon, Cory M and Doppa, Janardhan Rao},
  journal={Molecular Systems Design \& Engineering},
  volume={6},
  number={12},
  pages={1066--1086},
  year={2021},
  publisher={Royal Society of Chemistry}
}

@article{honarmandi2022accelerated,
  title={Accelerated materials design using batch Bayesian optimization: A case study for solving the inverse problem from materials microstructure to process specification},
  author={Honarmandi, P and Attari, V and Arroyave, R},
  journal={Computational Materials Science},
  volume={210},
  pages={111417},
  year={2022},
  publisher={Elsevier}
}

@article{smith2018less,
  title={Less is more: Sampling chemical space with active learning},
  author={Smith, Justin S and Nebgen, Ben and Lubbers, Nicholas and Isayev, Olexandr and Roitberg, Adrian E},
  journal={The Journal of chemical physics},
  volume={148},
  number={24},
  year={2018},
  publisher={AIP Publishing}
}

@article{zhong2022enlarging,
  title={Enlarging applicability domain of quantitative structure--activity relationship models through uncertainty-based active learning},
  author={Zhong, Shifa and Lambeth, Dylan R and Igou, Thomas K and Chen, Yongsheng},
  journal={ACS ES\&T Engineering},
  volume={2},
  number={7},
  pages={1211--1220},
  year={2022},
  publisher={ACS Publications}
}

@article{qi2024robust,
  title={Robust training of machine learning interatomic potentials with dimensionality reduction and stratified sampling},
  author={Qi, Ji and Ko, Tsz Wai and Wood, Brandon C and Pham, Tuan Anh and Ong, Shyue Ping},
  journal={npj Computational Materials},
  volume={10},
  number={1},
  pages={43},
  year={2024},
  publisher={Nature Publishing Group UK London}
}

@article{konyushkova2017learning,
  title={Learning active learning from data},
  author={Konyushkova, Ksenia and Sznitman, Raphael and Fua, Pascal},
  journal={Advances in neural information processing systems},
  volume={30},
  year={2017}
}

@article{angluin1988queries,
  title={Queries and concept learning},
  author={Angluin, Dana},
  journal={Machine learning},
  volume={2},
  pages={319--342},
  year={1988},
  publisher={Springer}
}

@article{atlas1989training,
  title={Training connectionist networks with queries and selective sampling},
  author={Atlas, Les and Cohn, David and Ladner, Richard},
  journal={Advances in neural information processing systems},
  volume={2},
  year={1989}
}

@inproceedings{lewis1995sequential,
  title={A sequential algorithm for training text classifiers: Corrigendum and additional data},
  author={Lewis, David D},
  booktitle={Acm Sigir Forum},
  volume={29},
  number={2},
  pages={13--19},
  year={1995},
  organization={ACM New York, NY, USA}
}

@inproceedings{murray2022addressing,
  title={Addressing bias in active learning with depth uncertainty networks... or not},
  author={Murray, Chelsea and Allingham, James U and Antor{\'a}n, Javier and Hern{\'a}ndez-Lobato, Jos{\'e} Miguel},
  booktitle={I (Still) Can't Believe It's Not Better! Workshop at NeurIPS 2021},
  pages={59--63},
  year={2022},
  organization={PMLR}
}

@article{yuan2018accelerated,
  title={Accelerated discovery of large electrostrains in BaTiO3-based piezoelectrics using active learning},
  author={Yuan, Ruihao and Liu, Zhen and Balachandran, Prasanna V and Xue, Deqing and Zhou, Yumei and Ding, Xiangdong and Sun, Jun and Xue, Dezhen and Lookman, Turab},
  journal={Advanced materials},
  volume={30},
  number={7},
  pages={1702884},
  year={2018},
  publisher={Wiley Online Library}
}

@article{pedersen2021bayesian,
  title={Bayesian optimization of high-entropy alloy compositions for electrocatalytic oxygen reduction},
  author={Pedersen, Jack K and Clausen, Christian M and Krysiak, Olga A and Xiao, Bin and Batchelor, Thomas AA and L{\"o}ffler, Tobias and Mints, Vladislav A and Banko, Lars and Arenz, Matthias and Savan, Alan and others},
  journal={Angewandte Chemie},
  volume={133},
  number={45},
  pages={24346--24354},
  year={2021},
  publisher={Wiley Online Library}
}

@article{kirklin2015open,
  title={The Open Quantum Materials Database (OQMD): assessing the accuracy of DFT formation energies},
  author={Kirklin, Scott and Saal, James E and Meredig, Bryce and Thompson, Alex and Doak, Jeff W and Aykol, Muratahan and R{\"u}hl, Stephan and Wolverton, Chris},
  journal={npj Computational Materials},
  volume={1},
  number={1},
  pages={1--15},
  year={2015},
  publisher={Nature Publishing Group}
}

@article{settles2009active,
  title={Active learning literature survey},
  author={Settles, Burr},
  year={2009},
  journal={arXiv preprint arXiv:1010.1010},
  publisher={University of Wisconsin-Madison Department of Computer Sciences}
}

@article{bartel2020critical,
  title={A critical examination of compound stability predictions from machine-learned formation energies},
  author={Bartel, Christopher J and Trewartha, Amalie and Wang, Qi and Dunn, Alexander and Jain, Anubhav and Ceder, Gerbrand},
  journal={npj computational materials},
  volume={6},
  number={1},
  pages={97},
  year={2020},
  publisher={Nature Publishing Group UK London}
}

@article{mackay1992information,
  title={Information-based objective functions for active data selection},
  author={MacKay, David JC},
  journal={Neural computation},
  volume={4},
  number={4},
  pages={590--604},
  year={1992},
  publisher={MIT Press One Rogers Street, Cambridge, MA 02142-1209, USA journals-info~…}
}

@article{dasgupta2011two,
  title={Two faces of active learning},
  author={Dasgupta, Sanjoy},
  journal={Theoretical computer science},
  volume={412},
  number={19},
  pages={1767--1781},
  year={2011},
  publisher={Elsevier}
}

@article{vasudevan2021machine,
  title={Machine learning for materials design and discovery},
  author={Vasudevan, Rama and Pilania, Ghanshyam and Balachandran, Prasanna V},
  journal={Journal of Applied Physics},
  volume={129},
  number={7},
  year={2021},
  publisher={AIP Publishing}
}

@article{batatia2025foundation,
  title={A foundation model for atomistic materials chemistry},
  author={Batatia, Ilyes and Benner, Philipp and Chiang, Yuan and Elena, Alin M and Kov{\'a}cs, D{\'a}vid P and Riebesell, Janosh and Advincula, Xavier R and Asta, Mark and Avaylon, Matthew and Baldwin, William J and others},
  journal={The Journal of chemical physics},
  volume={163},
  number={18},
  year={2025},
  publisher={AIP Publishing}
}

@article{axelrod2022learning,
  title={Learning matter: Materials design with machine learning and atomistic simulations},
  author={Axelrod, Simon and Schwalbe-Koda, Daniel and Mohapatra, Somesh and Damewood, James and Greenman, Kevin P and G{\'o}mez-Bombarelli, Rafael},
  journal={Accounts of Materials Research},
  volume={3},
  number={3},
  pages={343--357},
  year={2022},
  publisher={ACS Publications}
}

@article{farquhar2021statistical,
  title={On statistical bias in active learning: How and when to fix it},
  author={Farquhar, Sebastian and Gal, Yarin and Rainforth, Tom},
  journal={arXiv preprint arXiv:2101.11665},
  year={2021}
}

@article{li2024md,
  title={MD-HIT: Machine learning for material property prediction with dataset redundancy control},
  author={Li, Qin and Fu, Nihang and Omee, Sadman Sadeed and Hu, Jianjun},
  journal={npj Computational Materials},
  volume={10},
  number={1},
  pages={245},
  year={2024},
  publisher={Nature Publishing Group UK London}
}

@article{shahriari2015taking,
  title={Taking the human out of the loop: A review of Bayesian optimization},
  author={Shahriari, Bobak and Swersky, Kevin and Wang, Ziyu and Adams, Ryan P and De Freitas, Nando},
  journal={Proceedings of the IEEE},
  volume={104},
  number={1},
  pages={148--175},
  year={2015},
  publisher={IEEE}
}

@article{lim2021extrapolative,
  title={Extrapolative Bayesian optimization with Gaussian process and neural network ensemble surrogate models},
  author={Lim, Yee-Fun and Ng, Chee Koon and Vaitesswar, US and Hippalgaonkar, Kedar},
  journal={Advanced Intelligent Systems},
  volume={3},
  number={11},
  pages={2100101},
  year={2021},
  publisher={Wiley Online Library}
}

@article{moriconi2020high,
  title={High-dimensional Bayesian optimization using low-dimensional feature spaces},
  author={Moriconi, Riccardo and Deisenroth, Marc Peter and Sesh Kumar, KS},
  journal={Machine Learning},
  volume={109},
  pages={1925--1943},
  year={2020},
  publisher={Springer}
}

@article{varivoda2023materials,
  title={Materials property prediction with uncertainty quantification: A benchmark study},
  author={Varivoda, Daniel and Dong, Rongzhi and Omee, Sadman Sadeed and Hu, Jianjun},
  journal={Applied Physics Reviews},
  volume={10},
  number={2},
  year={2023},
  publisher={AIP Publishing}
}

@article{palmer2022calibration,
  title={Calibration after bootstrap for accurate uncertainty quantification in regression models},
  author={Palmer, Glenn and Du, Siqi and Politowicz, Alexander and Emory, Joshua Paul and Yang, Xiyu and Gautam, Anupraas and Gupta, Grishma and Li, Zhelong and Jacobs, Ryan and Morgan, Dane},
  journal={npj Computational Materials},
  volume={8},
  number={1},
  pages={115},
  year={2022},
  publisher={Nature Publishing Group UK London}
}

@article{chung2021uncertainty,
  title={Uncertainty toolbox: an open-source library for assessing, visualizing, and improving uncertainty quantification},
  author={Chung, Youngseog and Char, Ian and Guo, Han and Schneider, Jeff and Neiswanger, Willie},
  journal={arXiv preprint arXiv:2109.10254},
  year={2021}
}

@inproceedings{lataniotis2022uqlab,
  title={Uqlab 2.0 and uqcloud: open-source vs. cloud-based uncertainty quantification},
  author={Lataniotis, Christos and Marelli, Stefano and Sudret, Bruno},
  booktitle={SIAM Conference on Uncertainty Quantification (SIAM UQ 2022)},
  year={2022},
  organization={ETH Zurich, Institute of Structural Engineering}
}

@article{janet2019quantitative,
  title={A quantitative uncertainty metric controls error in neural network-driven chemical discovery},
  author={Janet, Jon Paul and Duan, Chenru and Yang, Tzuhsiung and Nandy, Aditya and Kulik, Heather J},
  journal={Chemical science},
  volume={10},
  number={34},
  pages={7913--7922},
  year={2019},
  publisher={Royal Society of Chemistry}
}

@article{hu2024realistic,
  title={Realistic material property prediction using domain adaptation based machine learning},
  author={Hu, Jeffrey and Liu, David and Fu, Nihang and Dong, Rongzhi},
  journal={Digital Discovery},
  volume={3},
  number={2},
  pages={300--312},
  year={2024},
  publisher={Royal Society of Chemistry}
}

@article{jung2023machine,
  title={Machine-learning driven global optimization of surface adsorbate geometries},
  author={Jung, Hyunwook and Sauerland, Lena and Stocker, Sina and Reuter, Karsten and Margraf, Johannes T},
  journal={npj Computational Materials},
  volume={9},
  number={1},
  pages={114},
  year={2023},
  publisher={Nature Publishing Group UK London}
}

@article{moon2024active,
  title={Active learning guides discovery of a champion four-metal perovskite oxide for oxygen evolution electrocatalysis},
  author={Moon, Junseok and Beker, Wiktor and Siek, Marta and Kim, Jiheon and Lee, Hyeon Seok and Hyeon, Taeghwan and Grzybowski, Bartosz A},
  journal={Nature Materials},
  volume={23},
  number={1},
  pages={108--115},
  year={2024},
  publisher={Nature Publishing Group UK London}
}

@article{gkatsis2025density,
  title={Density-Aware Active Learning for Materials Discovery: A Case Study on Functionalized Nanoporous Materials},
  author={Gkatsis, Vassilis and Maratos, Petros and Rekatsinas, Christoforos and Giannakopoulos, George and Krokidas, Panagiotis},
  journal={Physical Chemistry Chemical Physics},
  year={2025},
  publisher={Royal Society of Chemistry}
}

@article{schultz2025general,
  title={A general approach for determining applicability domain of machine learning models},
  author={Schultz, Lane E and Wang, Yiqi and Jacobs, Ryan and Morgan, Dane},
  journal={npj Computational Materials},
  volume={11},
  number={1},
  pages={95},
  year={2025},
  publisher={Nature Publishing Group UK London}
}

@article{korolev2022universal,
  title={A universal similarity based approach for predictive uncertainty quantification in materials science},
  author={Korolev, Vadim and Nevolin, Iurii and Protsenko, Pavel},
  journal={Scientific Reports},
  volume={12},
  number={1},
  pages={14931},
  year={2022},
  publisher={Nature Publishing Group UK London}
}

@inproceedings{kim2022defense,
  title={In defense of core-set: A density-aware core-set selection for active learning},
  author={Kim, Yeachan and Shin, Bonggun},
  booktitle={Proceedings of the 28th ACM SIGKDD conference on knowledge discovery and data mining},
  pages={804--812},
  year={2022}
}

@article{schein2007active,
  title={Active learning for logistic regression: an evaluation},
  author={Schein, Andrew I and Ungar, Lyle H},
  journal={Machine Learning},
  volume={68},
  pages={235--265},
  year={2007},
  publisher={Springer}
}

@article{fuhg2021state,
  title={State-of-the-art and comparative review of adaptive sampling methods for kriging},
  author={Fuhg, Jan N and Fau, Am{\'e}lie and Nackenhorst, Udo},
  journal={Archives of Computational Methods in Engineering},
  volume={28},
  pages={2689--2747},
  year={2021},
  publisher={Springer}
}

@article{kim2020machine,
  title={Machine-learned metrics for predicting the likelihood of success in materials discovery},
  author={Kim, Yoolhee and Kim, Edward and Antono, Erin and Meredig, Bryce and Ling, Julia},
  journal={npj Computational Materials},
  volume={6},
  number={1},
  pages={131},
  year={2020},
  publisher={Nature Publishing Group UK London}
}

@article{borg2023quantifying,
  title={Quantifying the performance of machine learning models in materials discovery},
  author={Borg, Christopher KH and Muckley, Eric S and Nyby, Clara and Saal, James E and Ward, Logan and Mehta, Apurva and Meredig, Bryce},
  journal={Digital Discovery},
  volume={2},
  number={2},
  pages={327--338},
  year={2023},
  publisher={Royal Society of Chemistry}
}

@article{jacobs2023role,
  title={Role of multifidelity data in sequential active learning materials discovery campaigns: case study of electronic bandgap},
  author={Jacobs, Ryan and Goins, Philip E and Morgan, Dane},
  journal={Machine Learning: Science and Technology},
  volume={4},
  number={4},
  pages={045060},
  year={2023},
  publisher={IOP Publishing}
}

@article{pyzer2022accelerating,
  title={Accelerating materials discovery using artificial intelligence, high performance computing and robotics},
  author={Pyzer-Knapp, Edward O and Pitera, Jed W and Staar, Peter WJ and Takeda, Seiji and Laino, Teodoro and Sanders, Daniel P and Sexton, James and Smith, John R and Curioni, Alessandro},
  journal={npj Computational Materials},
  volume={8},
  number={1},
  pages={84},
  year={2022},
  publisher={Nature Publishing Group UK London}
}

@article{wu2024race,
  title={Race to the bottom: Bayesian optimisation for chemical problems},
  author={Wu, Yifan and Walsh, Aron and Ganose, Alex M},
  journal={Digital Discovery},
  volume={3},
  number={6},
  pages={1086--1100},
  year={2024},
  publisher={Royal Society of Chemistry}
}

@article{yang2018active,
  title={Active learning Kriging model combining with kernel-density-estimation-based importance sampling method for the estimation of low failure probability},
  author={Yang, Xufeng and Liu, Yongshou and Mi, Caiying and Wang, Xiangjin},
  journal={Journal of Mechanical Design},
  volume={140},
  number={5},
  pages={051402},
  year={2018},
  publisher={American Society of Mechanical Engineers}
}

@article{xiong2013active,
  title={Active learning of constraints for semi-supervised clustering},
  author={Xiong, Sicheng and Azimi, Javad and Fern, Xiaoli Z},
  journal={IEEE Transactions on Knowledge and Data Engineering},
  volume={26},
  number={1},
  pages={43--54},
  year={2013},
  publisher={IEEE}
}

@article{bartok2013representing,
  title={On representing chemical environments},
  author={Bart{\'o}k, Albert P and Kondor, Risi and Cs{\'a}nyi, G{\'a}bor},
  journal={Physical Review B—Condensed Matter and Materials Physics},
  volume={87},
  number={18},
  pages={184115},
  year={2013},
  publisher={APS}
}

@article{hase2018phoenics,
  title={Phoenics: a Bayesian optimizer for chemistry},
  author={Hase, Florian and Roch, Lo{\"\i}c M and Kreisbeck, Christoph and Aspuru-Guzik, Al{\'a}n},
  journal={ACS central science},
  volume={4},
  number={9},
  pages={1134--1145},
  year={2018},
  publisher={ACS Publications}
}

@article{fromer2025batched,
  title={Batched Bayesian Optimization by Maximizing the Probability of Including the Optimum},
  author={Fromer, Jenna and Wang, Runzhong and Manjrekar, Mrunali and Tripp, Austin and Hern{\'a}ndez-Lobato, Jos{\'e} Miguel and Coley, Connor W},
  journal={Journal of Chemical Information and Modeling},
  year={2025},
  publisher={ACS Publications}
}

@article{roch2018chemos,
  title={ChemOS: orchestrating autonomous experimentation},
  author={Roch, Lo{\"\i}c M and H{\"a}se, Florian and Kreisbeck, Christoph and Tamayo-Mendoza, Teresa and Yunker, Lars PE and Hein, Jason E and Aspuru-Guzik, Al{\'a}n},
  journal={Science Robotics},
  volume={3},
  number={19},
  pages={eaat5559},
  year={2018},
  publisher={American Association for the Advancement of Science}
}

@article{hase2021olympus,
  title={Olympus: a benchmarking framework for noisy optimization and experiment planning},
  author={H{\"a}se, Florian and Aldeghi, Matteo and Hickman, Riley J and Roch, Lo{\"\i}c M and Christensen, Melodie and Liles, Elena and Hein, Jason E and Aspuru-Guzik, Al{\'a}n},
  journal={Machine Learning: Science and Technology},
  volume={2},
  number={3},
  pages={035021},
  year={2021},
  publisher={IOP Publishing}
}

@article{kusne2020fly,
  title={On-the-fly closed-loop materials discovery via Bayesian active learning},
  author={Kusne, A Gilad and Yu, Heshan and Wu, Changming and Zhang, Huairuo and Hattrick-Simpers, Jason and DeCost, Brian and Sarker, Suchismita and Oses, Corey and Toher, Cormac and Curtarolo, Stefano and others},
  journal={Nature communications},
  volume={11},
  number={1},
  pages={5966},
  year={2020},
  publisher={Nature Publishing Group UK London}
}

@article{yuan2024active,
  title={Active learning-based guided synthesis of engineered biochar for CO2 capture},
  author={Yuan, Xiangzhou and Suvarna, Manu and Lim, Juin Yau and P{\'e}rez-Ram{\'\i}rez, Javier and Wang, Xiaonan and Ok, Yong Sik},
  journal={Environmental Science \& Technology},
  volume={58},
  number={15},
  pages={6628--6636},
  year={2024},
  publisher={ACS Publications}
}

@article{min2020accelerated,
  title={Accelerated discovery of novel inorganic materials with desired properties using active learning},
  author={Min, Kyoungmin and Cho, Eunseog},
  journal={The Journal of Physical Chemistry C},
  volume={124},
  number={27},
  pages={14759--14767},
  year={2020},
  publisher={ACS Publications}
}

@article{sheng2020active,
  title={Active learning for the power factor prediction in diamond-like thermoelectric materials},
  author={Sheng, Ye and Wu, Yasong and Yang, Jiong and Lu, Wencong and Villars, Pierre and Zhang, Wenqing},
  journal={npj Computational Materials},
  volume={6},
  number={1},
  pages={171},
  year={2020},
  publisher={Nature Publishing Group UK London}
}

@article{fronzi2021active,
  title={Active learning in Bayesian neural networks for bandgap predictions of novel van der Waals heterostructures},
  author={Fronzi, Marco and Isayev, Olexandr and Winkler, David A and Shapter, Joseph G and Ellis, Amanda V and Sherrell, Peter C and Shepelin, Nick A and Corletto, Alexander and Ford, Michael J},
  journal={Advanced Intelligent Systems},
  volume={3},
  number={11},
  pages={2100080},
  year={2021},
  publisher={Wiley Online Library}
}

@article{tian2020role,
  title={Role of uncertainty estimation in accelerating materials development via active learning},
  author={Tian, Yuan and Yuan, Ruihao and Xue, Dezhen and Zhou, Yumei and Ding, Xiangdong and Sun, Jun and Lookman, Turab},
  journal={Journal of Applied Physics},
  volume={128},
  number={1},
  year={2020},
  publisher={AIP Publishing}
}

@article{hessmann2025accelerating,
  title={Accelerating crystal structure search through active learning with neural networks for rapid relaxations},
  author={Hessmann, Stefaan SP and Sch{\"u}tt, Kristof T and Gebauer, Niklas WA and Gastegger, Michael and Oguchi, Tamio and Yamashita, Tomoki},
  journal={npj Computational Materials},
  volume={11},
  number={1},
  pages={44},
  year={2025},
  publisher={Nature Publishing Group UK London}
}

@article{zhang2019active,
  title={Active learning of uniformly accurate interatomic potentials for materials simulation},
  author={Zhang, Linfeng and Lin, De-Ye and Wang, Han and Car, Roberto and E, Weinan},
  journal={Physical Review Materials},
  volume={3},
  number={2},
  pages={023804},
  year={2019},
  publisher={APS}
}

@article{nair2025materials,
  title={Materials-discovery workflow guided by symbolic regression for identifying acid-stable oxides for electrocatalysis},
  author={Nair, Akhil S and Foppa, Lucas and Scheffler, Matthias},
  journal={npj Computational Materials},
  volume={11},
  number={1},
  pages={1--7},
  year={2025},
  publisher={Nature Publishing Group}
}

@article{liu2024active,
  title={Active learning for regression of structure--property mapping: the importance of sampling and representation},
  author={Liu, Hao and Yucel, Berkay and Ganapathysubramanian, Baskar and Kalidindi, Surya R and Wheeler, Daniel and Wodo, Olga},
  journal={Digital Discovery},
  volume={3},
  number={10},
  pages={1997--2009},
  year={2024},
  publisher={Royal Society of Chemistry}
}

@article{pernot2023calibration,
  title={Calibration in machine learning uncertainty quantification: beyond consistency to target adaptivity},
  author={Pernot, Pascal},
  journal={APL Machine Learning},
  volume={1},
  number={4},
  year={2023},
  publisher={AIP Publishing}
}

@article{hwang2024overcoming,
  title={Overcoming overconfidence for active learning},
  author={Hwang, Yujin and Jo, Won and Hong, Juyoung and Choi, Yukyung},
  journal={IEEE Access},
  year={2024},
  publisher={IEEE}
}

@article{thomas2023calibration,
  title={Calibration of uncertainty in the active learning of machine learning force fields},
  author={Thomas-Mitchell, Adam and Hawe, Glenn and Popelier, Paul LA},
  journal={Machine Learning: Science and Technology},
  volume={4},
  number={4},
  pages={045034},
  year={2023},
  publisher={IOP Publishing}
}

@article{musielewicz2024improved,
  title={Improved Uncertainty Estimation of Graph Neural Network Potentials Using Engineered Latent Space Distances},
  author={Musielewicz, Joseph and Lan, Janice and Uyttendaele, Matt and Kitchin, John R},
  journal={The Journal of Physical Chemistry C},
  volume={128},
  number={49},
  pages={20799--20810},
  year={2024},
  publisher={ACS Publications}
}

@article{kavalsky2023much,
  title={By how much can closed-loop frameworks accelerate computational materials discovery?},
  author={Kavalsky, Lance and Hegde, Vinay I and Muckley, Eric and Johnson, Matthew S and Meredig, Bryce and Viswanathan, Venkatasubramanian},
  journal={Digital Discovery},
  volume={2},
  number={4},
  pages={1112--1125},
  year={2023},
  publisher={Royal Society of Chemistry}
}

@article{kavalsky2024multiobjective,
  title={A multiobjective closed-loop approach towards autonomous discovery of electrocatalysts for nitrogen reduction},
  author={Kavalsky, Lance and Hegde, Vinay I and Meredig, Bryce and Viswanathan, Venkatasubramanian},
  journal={Digital discovery},
  volume={3},
  number={5},
  pages={999--1010},
  year={2024},
  publisher={Royal Society of Chemistry}
}

@article{palizhati2022agents,
  title={Agents for sequential learning using multiple-fidelity data},
  author={Palizhati, Aini and Torrisi, Steven B and Aykol, Muratahan and Suram, Santosh K and Hummelsh{\o}j, Jens S and Montoya, Joseph H},
  journal={Scientific reports},
  volume={12},
  number={1},
  pages={4694},
  year={2022},
  publisher={Nature Publishing Group UK London}
}

@article{bi2025comprehensive,
  title={A comprehensive benchmark of active learning strategies with AutoML for small-sample regression in materials science},
  author={Bi, Jinghou and Xu, Yuanhao and Conrad, Felix and Wiemer, Hajo and Ihlenfeldt, Steffen},
  journal={Scientific Reports},
  volume={15},
  number={1},
  pages={37167},
  year={2025},
  publisher={Nature Publishing Group UK London}
}

@article{jose2024regression,
  title={Regression tree-based active learning},
  author={Jose, Ashna and de Mendon{\c{c}}a, Jo{\~a}o Paulo Almeida and Devijver, Emilie and Jakse, No{\"e}l and Monbet, Val{\'e}rie and Poloni, Roberta},
  journal={Data Mining and Knowledge Discovery},
  volume={38},
  number={2},
  pages={420--460},
  year={2024},
  publisher={Springer}
}

@article{sabanza2024best,
  title={Best practices for multi-fidelity bayesian optimization in materials and molecular research},
  author={Sabanza-Gil, V{\'\i}ctor and Barbano, Riccardo and Guti{\'e}rrez, Daniel Pacheco and Luterbacher, Jeremy S and Hern{\'a}ndez-Lobato, Jos{\'e} Miguel and Schwaller, Philippe and Roch, Lo{\"\i}c},
  journal={arXiv preprint arXiv:2410.00544},
  year={2024}
}

@article{xu2025multi,
  title={Multi-objective optimization in machine learning assisted materials design and discovery},
  author={Xu, Pengcheng and Ma, Yingying and Lu, Wencong and Li, Minjie and Zhao, Wenyue and Dai, Zhilong},
  journal={Journal of Materials Informatics},
  volume={5},
  number={2},
  pages={N--A},
  year={2025},
  publisher={OAE Publishing Inc.}
}

@article{trinquet2025optical,
  title={Optical materials discovery and design with federated databases and machine learning},
  author={Trinquet, Victor and Evans, Matthew L and Hargreaves, Cameron J and De Breuck, Pierre-Paul and Rignanese, Gian-Marco},
  journal={Faraday Discussions},
  volume={256},
  pages={459--482},
  year={2025},
  publisher={Royal Society of Chemistry}
}

@article{omee2024structure,
  title={Structure-based out-of-distribution (OOD) materials property prediction: a benchmark study},
  author={Omee, Sadman Sadeed and Fu, Nihang and Dong, Rongzhi and Hu, Ming and Hu, Jianjun},
  journal={npj Computational Materials},
  volume={10},
  number={1},
  pages={144},
  year={2024},
  publisher={Nature Publishing Group UK London}
}

@article{ouyang2018sisso,
  title={SISSO: A compressed-sensing method for identifying the best low-dimensional descriptor in an immensity of offered candidates},
  author={Ouyang, Runhai and Curtarolo, Stefano and Ahmetcik, Emre and Scheffler, Matthias and Ghiringhelli, Luca M},
  journal={Physical Review Materials},
  volume={2},
  number={8},
  pages={083802},
  year={2018},
  publisher={APS}
}

@article{xin2021active,
  title={Active-learning-based generative design for the discovery of wide-band-gap materials},
  author={Xin, Rui and Siriwardane, Edirisuriya MD and Song, Yuqi and Zhao, Yong and Louis, Steph-Yves and Nasiri, Alireza and Hu, Jianjun},
  journal={The Journal of Physical Chemistry C},
  volume={125},
  number={29},
  pages={16118--16128},
  year={2021},
  publisher={ACS Publications}
}

@inproceedings{schwarting2025steering,
  title={Steering an Active Learning Workflow Towards Novel Materials Discovery via Queue Prioritization},
  author={Schwarting, Marcus and Ward, Logan and Hudson, Nathaniel and Yan, Xiaoli and Blaiszik, Ben and Chaudhuri, Santanu and Huerta, Eliu and Foster, Ian},
  booktitle={2025 IEEE International Conference on eScience (eScience)},
  pages={30--38},
  year={2025},
  organization={IEEE}
}

@article{xian2025unlocking,
  title={Unlocking the black box beyond Bayesian global optimization for materials design using reinforcement learning},
  author={Xian, Yuehui and Ding, Xiangdong and Jiang, Xue and Zhou, Yumei and Sun, Jun and Xue, Dezhen and Lookman, Turab},
  journal={npj Computational Materials},
  volume={11},
  number={1},
  pages={1--11},
  year={2025},
  publisher={Nature Publishing Group}
}
\end{document}